\documentclass{emulateapj}
\usepackage{natbib}\usepackage{times}

\begin{document}   

\title{THE SURVIVAL OF NUCLEI IN JETS ASSOCIATED WITH CORE-COLLAPSE SUPERNOVAE AND GAMMA-RAY BURSTS}

\author{Shunsaku Horiuchi, Kohta Murase} 
\affiliation{Center for Cosmology and Astro-Particle Physics, The Ohio State University, 191 West Woodruff Avenue, Columbus, OH 43210, USA}

\author{Kunihito Ioka}
\affiliation{KEK Theory Center and the Graduate University for Advanced Studies (Sokendai), Tsukuba 305-0801, Japan}

\author{Peter M{\'e}sz{\'a}ros}
\affiliation{Deptment of Astronomy and Astrophysics, Deptment of Physics and Center for Particle Astrophysics, 525 Davey
Laboratory, Pennsylvania State University, University Park, PA 16802, USA}

\date{\today}   

\begin{abstract}
Heavy nuclei such as nickel-56 are synthesized in a wide range of core-collapse supernovae (CCSN), including energetic supernovae associated with gamma-ray bursts (GRBs). Recent studies suggest that jet-like outflows are a common feature of CCSN. These outflows may entrain synthesized nuclei at launch or during propagation, and provide interesting multi-messenger signals including heavy ultra-high-energy cosmic rays. Here, we investigate the destruction processes of nuclei during crossing from the stellar material into the jet material via a cocoon, and during propagation after being successfully loaded into the jet. We find that nuclei can survive for a range of jet parameters because collisional cooling is faster than spallation. While canonical high-luminosity GRB jets may contain nuclei, magnetic-dominated models or low-luminosity jets with small bulk Lorentz factors are more favorable for having a significant heavy nuclei component.
\end{abstract}   


\maketitle       

\section{Introduction} 

Recent studies have led to a canonical picture where long gamma-ray bursts (GRBs) are rare types of core-collapse supernovae (CCSN) that are accompanied by the launch of energetic relativistic jets~\citep[see, e.g.,][for reviews]{M06,WB06}. While the GRB is a rare phenomenon, a significantly larger fraction of CCSN could produce jets that do not successfully produce GRBs either because of energetic or collimation reasons. Many such jets may even be choked in their progenitor envelopes \citep{MW01}. Observationally, spectropolarimetry shows that the degree of asymmetry in CCSN increases with time and hence to greater depths of the CCSN, suggesting an association with the central engine and possible launch of bipolar jets \citep{WHHW01,CFLS10}. SN~1987A, the closest CCSN in modern times, shows a globally asymmetric expanding debris with an axis that roughly aligns with that of its rings \citep{WWH02}. Cas A, one of the well-known CCSN remnants in our Galaxy, may have been accompanied by an iron-rich jet component which may help to explain the observed apparent overturn of the Fe-rich ejecta \citep{WMC08,D10}. 

It is well known that nuclei can be synthesized not only in stellar nucleosynthesis but also during the CCSN, where pre-existing nuclei in the star are further fused into heavier nuclei via explosive nucleosynthesis. For CCSN associated with GRBs, such as SN~2003dh (GRB~030329) and SN~1998bw (GRB~980425), large kinetic energies of $\sim {10}^{52}$~erg are suggested and the inferred synthesized $^{56}$Ni masses are $\sim 0.5~M_{\odot}$ \citep{I98,WES99,WH03}. The large explosion energies may be due to a baryon-rich jet component driving the CCSN, which would lead to different elemental yields due to additional nucleosynthesis and/or efficient mixing and transport of stellar nuclei \citep{L02,PGF02,MNN02,MN03}. Alternatively, a large amount of $^{56}$Ni can be produced by a disk wind, where free nucleons ejected from the disk combine to form heavy nuclei \citep{MW99,PWH03,SM05}. The highly relativistic jet responsible for the GRB should have a smaller baryon mass, but may still pick-up and/or entrain stellar nuclei, nuclei synthesized by a wider jet, or nuclei synthesized by a disk wind. 

The fate of nuclei in relativistic jets is also interesting in light of recent reports on the nuclear composition of ultra-high-energy cosmic rays (UHECRs). The origin of UHECRs remains one of the great mysteries in high-energy astrophysics \citep[see, e.g.,][ for reviews]{H05,BEH09,BW09,KO11}. At the highest energies (above $\sim 50$~EeV), the extreme energies argue for extragalactic sources \citep[although Galactic transients may give a contribution around $\sim 10$ EeV, e.g.,][]{CKA10}. Source candidates fall into active galactic nuclei (AGNs), including radio-loud AGNs \citep{BS87,T90,NMA95}, radio-quiet AGNs \citep{PMM09}, young AGNs \citep{T90,TH11}, as well as powerful transient flares \citep{FG09,D09}; GRBs \citep[][]{W95,V95,MU95}, including sub-luminous GRBs accompanying relativistic ejecta in some form \citep{MINN06,WRM07,MINN08,C11,LW11}; the formation of rapidly rotating strongly magnetized protomagnetars \citep{A03,MMZ09,MGH11,K11,FKO12}; and galaxy cluster shocks \citep{NMA95,KRJ96,ISMA07}. Non-astrophysical sources include top-down scenarios where the UHECR arise from decays of massive relics from the early universe \citep[e.g.,][]{BKV97}. 

The composition of UHECRs is observationally inferred by measuring composition-dependent quantities of the showers made as UHECRs enter Earth's atmosphere. The Pierre Auger Observatory (PAO) finds that the average shower depth at shower maximum, $X_{\rm max}$, and its rms variation rms$(X_{\rm max})$, suggest the composition becoming increasingly dominated by heavy nuclei above the `ankle' of $10^{18.5}$ eV \citep[][]{Auger10a,AugerIRCR11}. Note that the PAO results have not been verified by the High Resolution Fly's Eye experiment \citep{HiRES10a} and the Telescope Array experiment \citep[e.g.,][]{TA11}, although the latter experiments observe a different hemisphere and their statistics are lower. In addition, these experiments do not confirm the directional anisotropies reported by PAO \citep{Auger07,Auger10b,HiRES08,HiRES10b}. Also, the PAO indicators have been shown to be inconsistent with each other \citep{SP12} and that they may not reflect a composition change given the current uncertainty in shower interaction physics \citep{WW11}. More observational and theoretical studies are required to settle these and other remaining issues.

If UHECRs are indeed composed of heavy nuclei, this provides important implications for UHECR sources. In the AGN and galaxy cluster shock origin scenarios, the matter being accelerated originates from the intergalactic medium, so that the mass fraction of Fe and heavier nuclei is small: $\sim 10^{-3}$ for solar metallicity. Although the dominance of nuclei in UHECRs may be explained by a rigidity-dependent acceleration mechanism, where the maximum Fe energy would be $Z_{\rm Fe} = 26$ times higher than those of protons, this requires the maximum accelerated energies in all contributing sources to be somewhat fine-tuned such that $E_{\rm p,max} \approx 10^{19}$ eV. Also, such a scenario predicts a larger proton-to-nuclei ratio at low energies than is actually observed \citep{LW09,Auger11}. An alternate scenario is that the environment supplying the injected particles is enriched in nuclei. The abundance of nuclei inferred from $X_{\rm max}$ and rms$(X_{\rm max})$ require high nuclei abundances \citep[see, e.g.,][and references therein]{A07,TAA11}. Massive-star origins, including GRBs, CCSNe with relativistic ejecta, and magnetars would be attractive sources in this sense, because as described above the environment contains large fractions of intermediate or heavy nuclei. Additionally, it has been shown that once loaded, nuclei may be accelerated to ultra-high energies and successfully survive in the dissipation regions of jets, including both classical high-luminosity GRBs (HL GRBs) and sub-luminous GRB \citep{MINN08,WRM08}. 

In this paper, we investigate the origins and survival of nuclei as the jet is still \textit{inside} the star. More specifically, we first discuss locations where nuclei may enter the jet medium and investigate whether nuclei can survive in each of these locations. Second, we investigate whether nuclei that have successfully made their way into the jet survive during the jet propagation through the progenitor star. These issues are different from previous works whose main focus were on the survival of UHECR nuclei in the emission regions (where the GRB occurs, typically outside the star). Note that UHECR acceleration is not expected inside the star even though our study would be useful for UHECR sources, and we mainly consider the survival of low-energy nuclei. 

The paper is organized as follows. In Section \ref{sec:sources}, we discuss sources of nuclei and discuss conditions for nuclei survival in each of them. In Section \ref{sec:survival}, we consider processes in the jet that may lead to nuclei disintegration and discuss conditions under which nuclei survive. Finally, we finish with discussions and conclusions in Section \ref{sec:conclusion}. We express quantities as $Q_x = Q/10^x$ in cgs units.

\section{Nuclei Origins and Survival} \label{sec:sources}

We first discuss three potential sources of nuclei in the context of GRBs. First, loading at the jet base, where a small amount of baryons may be loaded into the jet (the jet is assumed to be radiation or magnetically dominated). Depending on the jet parameters, nuclei survive or are disintegrated into free nucleons. Second, nucleosynthesis in the outflow itself, which occurs if nuclei are disintegrated into free nucleons at initial loading. Finally, entrainment, which is the loading of surrounding nuclei into the jet during jet propagation. 

\subsection{Nucleus Loading at the Jet Base}\label{sec:baseloading}

The progenitors of GRBs are massive stars with Fe cores of mass approximately $1$--$2 M_\odot$ that extend to radii a few $\times 10^8$ cm. It is thought that a few seconds after the onset of collapse, a compact object (either a neutron star or a black hole) forms at the center, surrounded by an equatorial disk supported by centrifugal forces. By some currently unconfirmed mechanism, an energy reservoir is tapped and energy is released near the collapsing core. In the ``collapsar'' model, the energy derives from the gravitational energy of rapid accretion, and neutrinos play the role of energy transport \citep{W93,MW99}. In this case, the energy is deposited dominantly as radiation energy. Alternatively, the jet may be dominated by a magnetic component, as for example in scenarios where the jet magnetic field threads the black hole event horizon and the power derives from the black hole spin energy \citep{BZ77}, where the jet magnetic field threads the surface of the accretion disk \citep{BP82,P03}, or where the power derives from the spin-down energy loss of a central magnetic protoneutron star \citep{U92}. 

We consider a jet with baryonic matter injected at rate $\dot{M}_0$, radiation energy at rate $L_{\rm rad,0} \gg \dot{M}_0 c^2$, and magnetic energy at rate $L_{\rm mag,0}$ (we will use collimation-corrected values, not the isotropic equivalent) at a radius $r_0$. The jet subsequently adiabatically expands as its radiation and/or magnetic energies are gradually converted to bulk kinetic energy. The maximum bulk Lorentz factor is set by $\eta = L_{\rm ke}/ (\dot{M} c^2)$, where $L_{\rm ke} = L_{\rm rad,0} + L_{\rm mag,0} $ is the kinetic luminosity after the end of the jet bulk acceleration phase\footnote{We caution that in magnetic-dominated jets such definitions may not strictly hold because of the model-dependent conversion of magnetic energy to kinetic energy; see Section \ref{sec:cocoon}.}. We defined a ``fireball'' as a jet where radiation dominates, i.e., $ L_{\rm rad,0} \gg L_{\rm mag,0}$, and a magnetic outflow as the opposite with $L_{\rm mag,0} \gg L_{\rm rad,0}$.

\begin{figure}[tb]
\includegraphics[width=3.25in]{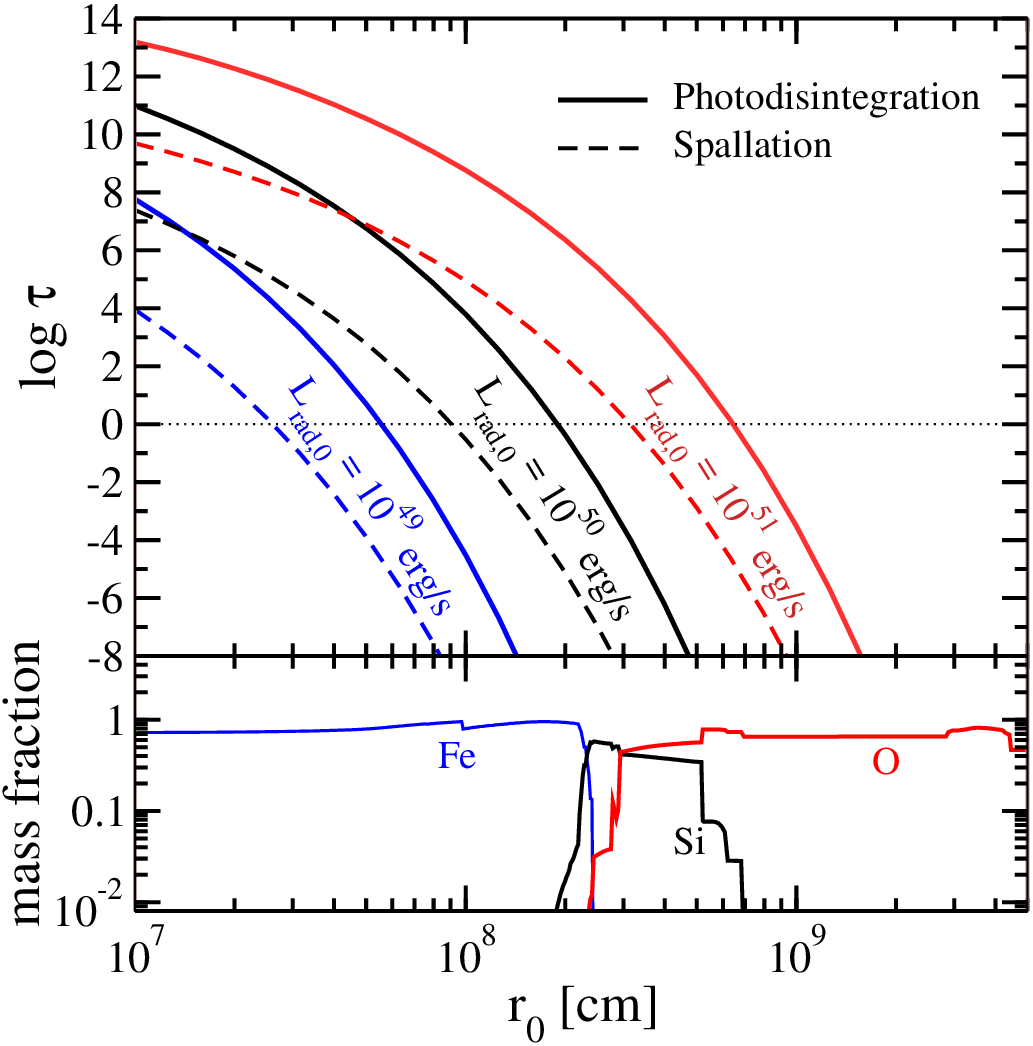}
\caption{\label{fig:injection} Optical depths for photodisintegration (solid) and spallation (dashed) as functions of jet injection radius, for various jet radiation luminosities. For a pure radiation fireball, $L_{\rm rad,0}$ corresponds to $L_{\rm ke}$ and the GRB luminosity by Equation (\ref{eq:GRBprompt}). In the case of a magnetic-dominated outflow, $ L_{\rm rad,0}$ may be significantly smaller for a given observed GRB luminosity, and nuclei loading would be easier. Bottom panel shows the dominant pre-supernova stellar abundances taken from the rotating $20 M_\odot$ pre-supernova progenitor model E20 of \cite{HLW00}. The abundances are illustrative: depending on the timing of jet launch, nuclei at small radii will be disintegrated by a supernova shock or synthesized by explosive nucleosynthesis before the jet propagates through (see the text). \\}
\end{figure}

The large values of $\eta \gtrsim 100$ required in GRB \citep{R75} necessitate a jet with small baryon loading. The quantitative predictions about where and how baryon loading occurs are tentative at best, but any baryons must come from the surrounding material. The composition of the pre-supernova stellar core is dominantly nuclei (bottom panel of Figure \ref{fig:injection}). At small radii this can be significantly altered by the collapse. For example, if a supernova shock is launched prior to the jet, nuclei out to a distance of $\sim 10^7$--$10^8$ cm would be disintegrated by the supernova shock. At later times, explosive nucleosynthesis can alter the composition. For example, \cite{FYH06} demonstrate with one-dimensional simulations that up to $\sim 1 M_\odot$ of $^{56}$Ni may be synthesized, depending on whether the core collapse proceeds to a black hole directly or via fall back, and also on the explosion energy. The detailed composition of the surroundings therefore depends on the GRB scenario and timing of jet launch relative to core collapse. In either case, it is expected that there are abundant nuclei especially for radii larger than $\sim 10^7$ cm. 

\begin{figure}[t]
\includegraphics[width=3.25in]{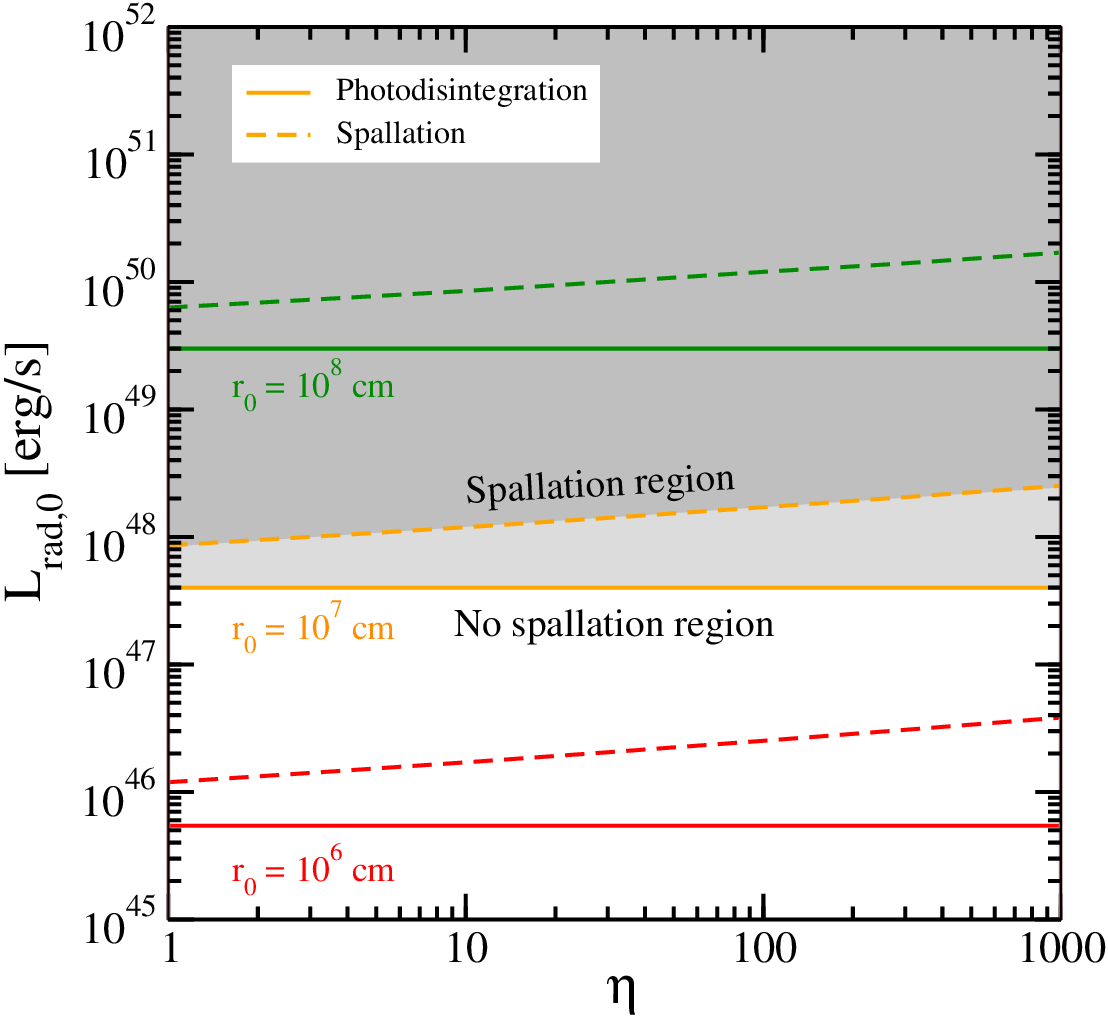}
\caption{\label{fig:result} Constraints on jet parameters for Fe nuclei survival during loading at the jet base. Photodisintegration (solid) and spallation (dashed) are shown separately for several values of $r_0$. Shaded regions result in nuclei destruction, while white areas allow nuclei survival (shades shown for our nominal $r_0 = 10^7$ cm). For magnetic-dominated jets, the situation is more favorable and quantitatively depends on the jet energy composition (see the text).}
\end{figure}

Any nuclei that are loaded into the jet are destroyed if it collides with particles or photons with energies exceeding the nuclear binding energy $\sim 10$ MeV in the nuclei rest frame. We consider a jet of luminosity $L_{\rm rad,0}$ injected at a radius $r_0$ with initial Lorentz factor $\Gamma_0 =1$. Its temperature is related to the radiation energy density, 
\begin{equation}
a T_0^4 = \frac{L_{\rm rad,0} }{ \Sigma_0 \Gamma_0^2 c},
\end{equation}
where $a$ is the radiation constant and $\Sigma_0 =  \Omega_0 {r_0}^2$ is the jet cross section. Typically, the jet is collimated already at injection due to rotation or magnetic geometries, and we adopt a solid angle $\Omega_0 = 0.1$ sr independent of $r_0$. This results in 
\begin{equation}
T_0 \approx 1.3 \, \Omega_{0,-1}^{-1/4} L_{\rm rad,50}^{1/4} r_{0,7}^{-1/2} \, {\rm MeV}.
\end{equation}
Therefore, nuclei will be destroyed by the high energy tail of the photon spectrum. The optical depth for photodisintegration is $\tau_{A\gamma} \simeq n_\gamma \sigma_{A\gamma} r_0 / \Gamma_0$, where $n_\gamma$ is the photon density and $\sigma_{A\gamma}$ is the photodisintegration cross section which peaks at approximately $10^{-25} \, {\rm cm^{2}}$ \citep[e.g.,][]{MINN08}. We require that $\tau_{A\gamma} < 1$ for the survival of nuclei during loading. Similarly for spallation we require that $\tau_{\rm sp} \simeq n_0 \sigma_{\rm sp} r_0 / \Gamma_0 < 1$, where $n_0$ is the comoving ion density,
\begin{equation} \label{eq:iondensity}
n_0 = \frac{L_{\rm ke} } {\Sigma_0 \eta \Gamma_0 \bar{A} m_p c^3},
\end{equation}
and $\bar{A}$ is the average mass number of the baryons in the jet. Also, $\sigma_{\rm sp} = \sigma_0 A^{2/3}$ with $\sigma_0 \approx 3 \times 10^{-26} \, {\rm cm^2}$ and $A=56$ is the mass number of the injected nuclei of interest (assumed Fe). 

In Figure \ref{fig:injection}, the photodisintegration and spallation optical depths are shown for three values of $L_{\rm rad,0}$. We adopt a thermal photon spectrum of temperature $T_0$, and assume that the jet ions have a Maxwell--Boltzmann velocity distribution with temperature equivalent to the photon temperature. Photodisintegration is more important than spallation, because the larger photon density compensates the slightly smaller peak photodisintegration cross section. Note that in Figure \ref{fig:injection} we adopt $\bar{A} = 1$, which gives the largest spallation target density possible. 

As expected, for a GRB fireball of typical luminosities ($L_{\rm rad,0} = 10^{49}$--$10^{51} \, {\rm erg\,s^{-1}}$), the high $T_0$ results in nuclei being photodisintegrated at loading. However, since photodisintegration occurs with the exponential tail of the photon spectrum, nuclei survival is highly sensitive on $T_0$ and hence GRB parameters. For example, nuclei survival is possible for $r_0$ greater than $10^8$--$10^9$ cm. Values beyond $10^8$ cm are somewhat large but plausible if the jet is powered by an accretion disk. However, in this case it may be difficult to achieve short variable time scales on the order of ms observed in some GRBs, unless they are imprinted by instabilities during jet propagation. Figure \ref{fig:injection} also demonstrates that nuclei survival is more easily possible for sub-luminous GRBs. We quantify these constraints in Figure \ref{fig:result}. 

It is easier to initially load nuclei in magnetic-dominated jets ($L_{\rm mag,0} \gg L_{\rm rad,0}$), because $L_{\rm rad,0}$ is smaller for a given observed GRB luminosity and $T_0$ is correspondingly smaller. Both photodisintegration and spallation become less effective.

\subsection{Nucleosynthesis by Outflows} \label{sec:nucleosynthesis}

The fireball is composed of free nucleons if the photodisintegration optical depth at jet launch is larger than unity.  As the jet expands and cools, nucleons combined to form nuclei. The freeze-out composition of HL GRB jets has been investigated by various authors, and strongly depends on the radiation energy content of the jet. For a canonical GRB fireball of initial temperature $\sim 1$ MeV, the entropy per baryon is necessarily high, $S \gtrsim 10^5 \, k_b$ nucleon$^{-1}$. Free nucleons combine to form $\alpha$ particles only once the deuterium bottleneck has been broken, which for higher entropies occurs at lower jet densities where the further processing into carbon and higher nuclei are not fast enough compared to the expansion time scale.  As a result, few elements heavier than He are formed, similar to big bang nucleosynthesis \citep{L02,PGF02,B03}. 

The situation is dramatically different for magnetically dominated jets, where most of the energy is stored in magnetic energy and the jet entropy is necessarily lower, $S \sim 10$--$300 \, k_b$ nucleon$^{-1}$.  Under such conditions $\alpha$ recombination occurs at higher densities such that heavier elements can be formed efficiently via the triple-$\alpha$ process and subsequent $\alpha$ captures.  Focusing on the magnetic jets of a millisecond protomagnetar central engine, \cite{MGH11} showed that the jet composition may indeed be dominated by heavy nuclei.  Also, depending on how neutron-rich the jet matter is \citep{MTQ08}, neutron-catalyzed $\alpha$ formation and neutron captures open the path for the rapid synthesis of Fe-peak and heavier $n$-capture nuclei (e.g., $A \gtrsim 90$). Similar nucleosynthesis is also realized in more baryon-rich jets with lower entropies \citep{I03}. Although the baryonic jets cannot produce the classical GRB phenomenon, it can be relevant for sub-luminous GRBs and hypernovae where $\eta$ can be much lower than $\eta \sim 100$--$1000$. 

When the central engine consists of a black hole and accretion disk system, hot outflows from the disk are expected, where the disk wind with $S \sim 10$--$100 \, k_b$ nucleon$^{-1}$ can account for the amount of $^{56}$Ni observed in hypernovae associated with GRBs \citep[e.g.,][]{SM05}. We discuss the entrainment of external nuclei next. 
 
\subsection{Nucleus Entrainment During Propagation}

As the jet propagates through the star, it can pick up baryons from the stellar core, surrounding wider jet, or the disk wind environment.  Since these environments can be nuclei-rich, we discuss how nuclei may survive during the entrainment process. We will assume a proton-dominated jet ($\bar{A}=1$) for this purpose.  For our canonical jet, we consider a classical high-luminosity GRB jet, fixing $\Gamma_0=1$ and $\Omega_0 = 0.1 \, {\rm sr}$. Further, we adopt $L_{\rm ke} = 10^{50} \, {\rm erg \, s^{-1}}$, $\eta = 100$, and $r_0 = 10^7 \, {\rm cm}$, and show dependencies where applicable. For illustration purposes, we adopt the rotating $20 M_\odot$ pre-supernova progenitor model E20 of \cite{HLW00}, whose density profile can be well approximated by $\rho_*(r) = 1.7 \times 10^5 (r/10^9 \, {\rm cm})^{-n} \, {\rm g \, cm^{-3}}$ with $n \approx 3$.

\subsubsection{Cocoon Properties} \label{sec:cocoon}

After the onset of collapse, the lack of centrifugal force along the rotational axis leads to free mass infall and the formation of a funnel region where conditions are favorable for the launch of a jet \citep{W93,MW99}. In magnetic-dominated jets, the outflow initially flows along open field lines. However, self-collimation via magnetic hoop stress fails in relativistic flows \citep[e.g.,][]{S85,B06}, and recent works show how interactions with the star may redirect the flow toward the poles and produce relativistic bipolar jets along the rotational axis \citep{B07,B08,KB07,UM07}. We model the jet cross section as
\begin{equation} \label{eq:crosssection}
\Sigma(r) = \Sigma_0 \left( \frac{r}{r_0} \right)^{\xi},
\end{equation}
where for example $\xi =2$ corresponds to a conical (or spherical symmetric) jet, and $\xi =1$ corresponds to a funnel-like jet \citep[e.g.,][]{MR01}. 

After injection, the GRB jet accelerates as its radiation or magnetic energy is converted to bulk kinetic energy. For a radiative fireball expanding adiabatically with cross section $\Sigma$, energy conservation yields \citep[see also, e.g.,][]{TISN07,I11}
\begin{equation} \label{eq:gamma}
\Gamma_j(r) = \left\{ 
\begin{array}{ll}
\Gamma_0 (r/r_0)^{\xi /2} & \quad r < r_{\rm sat} \\
\eta & \quad r > r_{\rm sat},
\end{array}\right.
\end{equation}
where $r_{\rm sat} = \eta^{2/\xi} r_0$ is the saturation radius where the terminal Lorentz factor is reached. Often $r< r_{\rm sat}$ is called the jet or bulk acceleration phase and $r>r_{\rm sat}$ is the jet coasting phase.

The relativistic jet slows down abruptly in a narrow layer at the head of the jet where it comes into contact with the overlaying stellar material. Due to the large ram pressure experienced, a strong reverse shock forms and the jet head is decelerated to a sub-relativistic velocity $\beta_h < 1$. Further upstream, a forward shock forms that shock heats the stellar material. The jet head velocity is set by balancing the ram pressures applied on the forward and reverse shocks. We use the analytic approximations of \cite{M03},
\begin{equation} \label{eq:headvelocity}
\beta_h = \beta_j \frac{1}{1+ \tilde{L}^{-1/2}},
\end{equation}
where $\tilde{L}$ is a dimensionless parameter defined as
\begin{equation}
\tilde{L} \simeq \frac{ L_{\rm ke} }{ \Sigma (r) \rho_* c^3}.
\end{equation}
In the limit $\tilde{L} \ll 1$, which corresponds to a relativistic reverse shock (as viewed in the jet frame) as is generally the case for jet propagation in the star, Equation (\ref{eq:headvelocity}) can be linearized, and 
\begin{equation} \label{eq:headvelocityLimit}
\beta_h  \simeq \tilde{L}^{1/2}  \propto {L_{\rm ke}}^{1/2} r^{(n-\xi)/2} r_0^{(\xi-2)/2}.
\end{equation}

We call the shocked stellar and shocked jet material collectively as the jet head. At first, the jet head pressure is smaller than the surrounding stellar pressure and the jet head remains pressure confined. The jet head pressure is $P_h \simeq U_h /3$, where $U_h \simeq 4 {\Gamma_j}^2 n_p m_p c^2$ is the jet head energy density and $n_p$ is the proton density, i.e., the ion density
\begin{equation} \label{eq:iondensity}
n_i(r) = \frac{L_{\rm ke} } {\Sigma(r) \eta \Gamma_j \bar{A} m_p c^3},
\end{equation}
with $\bar{A}=1$. Hence $P_h$ falls as $ \propto r^{- \xi/2}$ during the bulk acceleration phase and $ \propto r^{- \xi}$ during the coasting phase, while the stellar pressure falls as $P_* \propto {\rho_*}^{4/3} \propto r^{-4n/3}$. Therefore, the jet head pressure eventually overtakes the stellar pressure and starts to overflow. We call the radius at which the pressures become equal as $r_c$; cocoon formation occurs for $r>r_c$. For typical parameters, $r_c$ is in the range $10^8$--$10^9$ cm and cocoon formation occurs during the bulk acceleration phase \citep{MR01}. 

For magnetic-dominated outflows, the conversion of magnetic energy to kinetic energy is prolonged since only part of the magnetic luminosity gets converted directly into kinetic luminosity, the other part being converted to kinetic energy in a two-stage process through thermal energy \citep[e.g.,][]{MR11}. Furthermore, the conversion is not efficient in unconfined, time-stationary outflows in ideal MHD \citep[e.g.,][]{GJ70,BT99}. Therefore, models for full jet acceleration employing a combination of differential collimation, time variability, or violations of ideal MHD have been studied. These result in $\Gamma_j$ that increases roughly as $\propto r^{1/3}$ \citep[e.g.,][]{D02,GKS11,M11} or $\propto r^{1/2}$ \citep[e.g.,][]{MU12}. For illustration, we adopt the calculations of \cite{MU12} who find the jet dynamics well fit by $\Gamma_j \propto r^{1/2}$ and $\Sigma \propto r^{5/4}$ during jet propagation in the star (see their Figure 5). In this case the jet head pressure falls as $\propto r^{-3/4}$, so cocoon formation is again inevitable, occurring in the range $10^8$--$10^9$ cm. Note that our adopted magnetic-dominated jet has properties---such as particle density, energy density, bulk acceleration, and so on---in between our conical and funnel jets.

\subsubsection{Survival of Nuclei in theCcocoon}\label{sec:survival_cocoon}

The freshly formed cocoon consists of an inner region composed of shocked jet material and an outer region composed of shocked stellar material. The two are separated by a contact discontinuity. The contact discontinuity is dynamically unstable and studies find that it remains on the order of seconds \citep[e.g.,][]{MA09}. After that, the outer and inner cocoon material mix. We discuss how stellar nuclei may survive until this last phase of cocoon evolution. Note that the discussions below will also apply to magnetic outflows if they interact with the progenitor to produce a cocoon. 

Nuclei in the star ahead of the jet are first shocked by the forward shock approaching with velocity $\sim \beta_h$. Spallation occurs when the energy of a collision between nuclei and jet head protons exceeds the nuclear binding energy $ \approx 10$ MeV in the nuclei rest frame. Therefore, external nuclei survive if the jet head velocity is less than $\beta_{\rm sp} \approx 0.14$ in the stellar frame. The jet head bulk velocity for a conical ($\xi =2$) jet,
\begin{equation} \label{eq:headvelocityNum}
\beta_h^{(2)} \sim 0.01 \, L_{\rm ke,50}^{1/2} r_9^{1/2},
\end{equation}
typically exceeds $\beta_{\rm sp}$ for radii close to $10^{11}$ cm. For more collimated jets, $\beta_{\rm sp}$ is reached at smaller radii. For example, the jet head velocity of a $\xi = 1$ jet of the same jet parameters is ${\beta_h}^{(1)} \sim 0.1 {L_{\rm ke,50}}^{1/2} {r_9} {r_{0,7}}^{-1/2}$. Also note that $\beta_h$ grows faster with radius and the saturation radius is larger. Therefore, this channel for nuclei loading is limited to jet head radii less than $10^9$--$10^{11}$ cm depending on jet parameters. 

Nuclei can also enter the cocoon directly through its boundary with stellar matter. The cocoon is overpressured and expands into the star at velocity $\beta_c$ given by pressure equilibrium, $\rho_* c^2 {\beta_c}^2 = P_c$, where $P_c = E_c / (3 V_c)$ is the radiation dominated cocoon pressure and $E_c \simeq L_{\rm ke} (t-r/c)$ is the total energy deposited in the cocoon \citep[][]{BC89,M03}. We approximate the cocoon as a cylinder of height $r$ and base length $x_c \approx c \beta_c t$, so that its volume is $V_c = \pi c^2 {\beta_c}^2 t^2 r$. The cocoon expansion velocity is then
\begin{equation} \label{eq:cocoonvelocity}
\beta_c \simeq \left( \frac{P_c}{\rho_* c^2} \right)^{1/2} \propto L_{\rm ke}^{3/8} r^{(3n-\xi-4)/8} r_0^{(\xi-2)/8},
\end{equation}
which is significantly slower than the head velocity and depends weakly on jet parameters. For example, even for a $\xi =1$ jet with our nominal jet parameters (i.e., $L_{\rm ke} = 10^{50} \, {\rm erg \, s^{-1}}$, $\eta = 100$, and $r_0 = 10^7 \, {\rm cm}$), the cocoon expands at only ${\beta_c}^{(1)} \sim 0.04 \, {r_{10}}^{1/2} $. Furthermore, the flow rate of material into the cocoon through the cocoon boundary can be substantial. The flow rate scales as the product of the expansion velocity and surface area. For the jet head this is $\sim \pi {x_c}^2 c \beta_h $. For the cocoon boundary this is $\sim 2 \pi x_c r c \beta_c $ which can be rewritten $ \sim 2 \pi {x_c}^2 c \beta_h $ using the approximations $r \approx c \beta_h t$ and $x_c \approx c\beta_c t$. Thus, the flow rates through the jet head and cocoon boundary are likely at least comparable.

Once in the outer cocoon, nuclei must survive photodisintegration. The photon temperature in the bulk acceleration phase is
\begin{equation}
T_h \simeq \left(  \frac{\epsilon_e U_h }{a} \right)^{1/4} \propto \epsilon_e^{1/4} L_{\rm ke}^{1/4} r^{-\xi/8} \eta^{-1/4} r_0^{(\xi-4)/8},
\end{equation}
and for a conical ($\xi =2$) jet this yields
\begin{equation}
T_h^{(2)} \sim 100 \, \epsilon_{e,-1}^{1/4} L_{\rm ke,50}^{1/4} r_9^{-1/4} \eta_2^{-1/4} r_{0,7}^{-1/4}\, {\rm keV },
\end{equation}
where for generality we adopt a parameter $\epsilon_e \lesssim 1$ to characterize the fraction of the jet head internal energy that is radiated by electrons. Throughout, we adopt $\epsilon_e =  0.1$ and show dependencies on $\epsilon_e$. We quote results for the bulk acceleration phase because the GRB jet is typically in the acceleration phase when it advances through the stellar core, unless it is a baryon-rich jet with a small saturation radius. Note that for a $\xi=1$ jet of the same parameters, ${T_h}^{(1)} \sim 160$ keV. Thus temperatures are insufficient for rapid photodisintegration. Also, the cocoon would be turbulent, with flow speeds that may reach the sound velocity $\sim c/ \sqrt{3}$. Nuclei being carried by such flows will exceed the spallation energy threshold. As we detail in Section \ref{sec:entrainment}, only relativistic nuclei above a critical energy are spalled, because lower energy nuclei are expected to thermalize rapidly. 

Next, we must check that nuclei survive potential collisions with protons in the inner cocoon. In particular, jet protons may acquire relativistic random velocities at the reverse shock. The protons could be trapped in the fireball by strong magnetic fields that are either generated in situ or advected from the central engine, and at the reverse shock their directions could be isotropized so that they maintain their velocities comparable to $\Gamma_j$ in magnitude \citep[e.g., see discussions in][]{I10}. In this case, the relativistic protons can cause spallation reactions. 

However, the relativistic protons typically lose energy very rapidly. For example, relativistic protons lose energy by $\pi$-production at an energy loss rate
\begin{equation}
\nu_{pp\pi} \simeq 0.2 (4 \Gamma_j n_p) \sigma_{0} c  \propto L_{\rm ke} r^{- \xi} \eta^{-1} r_0^{\xi-2},
\end{equation} 
where the quantity in brackets is the jet head particle density and $\sigma_0$ is again $3 \times 10^{-26} \, {\rm cm^2}$. For our canonical jet, ${\nu_{pp\pi}}^{(2)} \sim 2 \times 10^{8} \, {r_9}^{-2} \, {\rm s^{-1}}$ and ${\nu_{pp\pi}}^{(1)} \sim 1 \times 10^{10} \, {r_9}^{-1} \, {\rm s^{-1}}$ for $\xi =2$ and $\xi =1$, respectively. $\pi$-production reduces the proton kinetic energy down to around $70$ MeV. Protons also lose energy by processes such as $e^\pm$-production and Coulomb interactions which are particularly important to reduce the proton kinetic energy further. Since electrons rapidly lose their energy by Compton scattering and Bremsstrahlung emission, the electrons thermalize and their temperature is lower than protons. Thus, $\sim 70$ MeV protons lose energy to electrons at a rate (see also Section \ref{sec:entrainment})
\begin{equation} \label{eq:pe}
\nu_{pe} \simeq \frac{32 \sqrt{\pi} n_e q^4 {\rm ln} \, \Lambda_{pe}}{3 m_p m_e v_e^3} \propto L_{\rm ke} r^{- \xi} \eta^{-1} r_0^{\xi-2},
\end{equation}
where ${\rm ln}\, \Lambda \sim 10$ is the Coulomb logarithm, $v_e$ is the electron velocity\footnote{We assume that $T_e$ is equivalent to the radiation temperature $T_h$ because of Compton scattering. When the temperature is $T_h \lesssim 10^5$ eV, $v_e \simeq (3 T_h/m_e)^{1/2}$ and the $\nu_{pe}$ dependency changes accordingly.}, and $n_e = n_p$ if the only electrons in the jet are those associated with the ions; if $e^\pm$ pair production in the jet head can increase the proton cooling rate (see Section \ref{sec:shocks}). For our canonical GRB jet, this yields ${\nu_{pe}}^{(2)} \sim 1 \times 10^{9} \, {r_9}^{- 2} \, {\rm s^{-1}}$ and ${\nu_{pe} }^{(1)} \sim 7 \times 10^{10} \, {r_9}^{-1} \, {\rm s^{-1}}$ for $\xi =2$ and $\xi =1$, respectively. 

The short cooling timescales imply that relativistic protons occupy only a very thin region around the reverse shock. The proton cooling time scale is also much shorter than the lifetime of the contact discontinuity separating the outer and inner cocoons. Thus, relativistic protons would have lost much of their energy by the time nuclei in the outer cocoon come into contact with jet protons in the inner cocoon. Note that this conclusion can break down for jets with extremely large $\eta$, where the optical depth to $pp$ collisions can be below unity and protons will not thermalize \citep{I10}.

Finally, nuclei must survive in the cocoon. The cocoon temperature is determined by the energy density of the cocoon. We estimate the cocoon temperature as
\begin{eqnarray} \label{eq:Tc}
T_c \simeq \left( \frac{\epsilon_e E_c}{a V_c} \right)^{1/4} \propto \epsilon_e^{1/4} L_{\rm ke}^{3/16} r^{-(n+\xi+4)/16} r_0^{(\xi-2)/16},
\end{eqnarray}
where $\epsilon_e = 0.1$ as before gives ${T_c}^{(2)} \sim 80 \, {r_9}^{-9/16}$ eV and ${T_c}^{(1)} \sim 100 \, {r_9}^{-1/2}$ keV for $\xi =2$ and $\xi =1$, respectively. These are too low for significant photodisintegration. 

In conclusion, external nuclei enter the cocoon through the jet head and the boundary between the cocoon and the star. Those entering through the jet head survive provided the jet head velocity is slower than $\beta_{\rm sp}$. This is usually satisfied for $r \lesssim 10^{11} \, {\rm cm}$, but for more penetrating jets can be as limited as $r \lesssim 10^9$ cm. Those entering through the cocoon boundary survive spallation. Once in the cocoon, nuclei survive both spallation and photodisintegration. 

\subsubsection{Survival of Nuclei During Entrainment} \label{sec:entrainment}

A classical HL GRB jet is likely to be highly relativistic at its central cross section, moving with bulk Lorentz factor $\Gamma_j$. Surrounding this is a transition layer where the velocity decreases from relativistic to non-relativistic values. Outside the transition layer lies the non-relativistic nuclei-rich cocoon. The growth time of shear-driven instabilities at the transition layer is much smaller than the duration of the jet, and the transition layer likely contains rapid fluctuations in thermodynamic quantities \citep[][]{AIMU02}. Here, we discuss whether nuclei are destroyed when nuclei cross into the jet through such transition layers. 

When a nucleus in the cocoon moves into the jet plasma, it has an extremely short thermal relaxation time corresponding to $\sim 1/(\nu_{Ap}+\nu_{Ae})$, where $\nu_{Ap}$ and $\nu_{Ae}$ are the energy loss rates on jet protons and electrons, respectively. These can be written as $\nu = 2 \nu_S - \nu_\perp - \nu_\parallel$, where $\nu_S$, $\nu_\perp$, and $ \nu_\parallel$ are the slowing down rate (or momentum loss rate), the pitch-angle diffusion rate, and the parallel velocity diffusion rate, respectively. The pitch angle and parallel velocity diffusion refer to the perpendicular and parallel velocity components of test particles spreading in velocity space through multiple Coulomb scatterings. For sufficiently energetic test particles, $\nu_\perp$ and $ \nu_\parallel$ are smaller than $\nu_S$; in other words, the overall slowing down of a test particle is more significant than its diffusion in velocity space. We may thus approximate $\nu \approx 2\nu_S$. The slowing down rate is defined $\nu_S =  - \langle \Delta v_\| \rangle / v $, where $v$ is the particle velocity and $v_\|$ is in the direction of the nucleus motion \citep[e.g.,][]{S56}, and
\begin{eqnarray} \label{eq:Ai}
\nu_{S,Ae} = \frac{(1+ \gamma_A m_A/m_e) A_D G(v_A / v_e)}{ v_A v_e^2},
\end{eqnarray}
where $v_A$ is the nuclei velocity and $\gamma_A$ is its Lorentz factor, $A_D$ and $G(y)$ are
\begin{eqnarray}
A_D &=& \frac{8 \pi n_e q^4 Z_A^2 Z_e^2 {\rm ln} \, \Lambda_{Ae} }{\gamma_A^2 m_A^2}, \\
G(y) &=& \frac{\Phi(y) - y \Phi'(y) }{ 2 y^2 },
\end{eqnarray}
and $\Phi(y)$ is the usual error function. This is the energy loss rate on jet electrons; the same expression with $e$ replaced by $p$ applies for energy loss on jet protons. The expression becomes inaccurate when $v_A / v_e \gtrsim {\rm ln} \, \Lambda_{Ae}$, because terms ignored in its derivation become important \citep{S56}. However, in our case the jet electron velocity is already close to $c$ and this is not a serious concern. Indeed, for large $y$ we can approximate $G(y) \to 1/(2y^2)$ and we find that $\nu$ agrees numerically with the well-documented energy loss rate of relativistic cosmic rays propagating through fully ionized plasmas \cite[see, e.g., Equation (5.3.40) in Section 5.3.8.1 of][]{S02}. Note that for $v_A < v_e$, $G(y) \approx (2y) / (3 \sqrt{\pi})$ and one obtains Equation (\ref{eq:pe}). For reference, the expressions for the pitch-angle diffusion and the parallel velocity diffusion rates are \citep[][]{S56}
\begin{eqnarray} 
\nu_{\perp,Ae} &=& \frac{A_D \{ \Phi(v_A/v_e) - G(v_A/v_e) \} }{v_A^3}, \\
\nu_{\parallel,Ae} &=& \frac{4 A_D G(v_A/v_e) }{v_A^3}.
\end{eqnarray}
These affect the energy loss rates at small nuclei energies.

\begin{figure}[tb]
\includegraphics[width=3.25in]{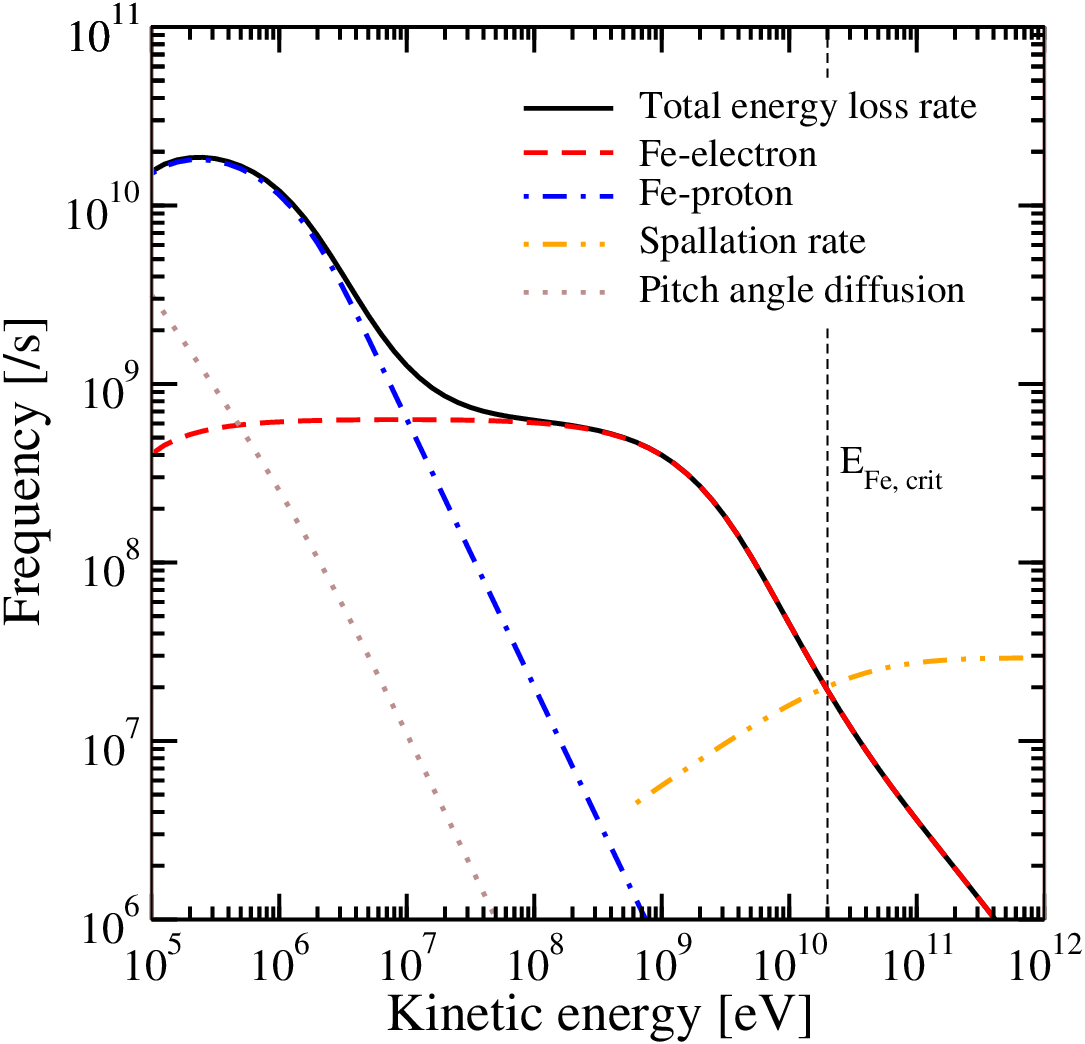}
\caption{\label{fig:timescales}Energy loss rates and spallation rate for an energetic test Fe nuclei moving through the jet plasma. Jet electrons and protons are assumed to have temperatures equivalent to the jet radiation temperature. Jet parameters are  $r_0 = 10^7 \, {\rm cm}$, $L_{\rm ke} = 10^{50} \, {\rm erg\,s^{-1}}$, $\eta=100$, and $r=10^9$ cm. Pure proton jet composition with $n_e = n_p$ has been assumed. Below $E_{\rm Fe, crit} \sim 20$ GeV, Fe nuclei lose energy before they are spalled, even though spallation is energetically allowed. \\ \\}
\end{figure}

In Figure \ref{fig:timescales}, we show the energy loss rate $\nu = 2 \nu_S - \nu_\perp - \nu_\parallel$ for a test Fe nuclei in the jet plasma at $r=10^9$ cm. The jet ion component is assumed to be proton dominated ($Z_i = 1$) and the electron and proton temperatures are taken to be equivalent to the jet radiation temperature; for the bulk acceleration phase this is $T_j \simeq T_0 (r/r_0)^{-\xi/2}$. For high-energy nuclei, energy loss on electrons is more important than energy loss on protons because of the faster electron velocities. We compare the energy loss rates to the spallation rate $\nu_{\rm sp} \simeq n_p \sigma_{sp} v_{\rm Fe}$. We do not show the spallation energy loss rate because we wish to remain conservative and assume that a single spallation event affects the composition. We see that at relativistic nuclei energies, spallation dominates over Coulomb cooling rates. However, below $E_{\rm Fe, crit} \sim 20$ GeV, Fe nuclei lose energy before they are spalled, even though spallation is energetically allowed. These nuclei lose energy to electrons on an exponential time scale $E_{\rm Fe} \propto e^{-\nu_{Ae, {\rm crit}} t}$, where $\nu_{Ae, {\rm crit}} = \nu_{Ae}(E_{\rm Fe, crit})$. 

Importantly, the critical energy $E_{\rm Fe, crit}$ depends very weakly on the radius and on GRB parameters. This is because in the limit that $v_A > v_e$, the Coulomb cooling rate can be approximated as 
\begin{equation}
\nu_{Ae} \simeq \frac{8 \pi n_e q^4 Z_A^2 Z_e^2 {\rm ln} \, \Lambda_{Ae}}{\gamma_A m_A m_e v_A^3} \propto n_p,
\end{equation}
for $Z_i =1$. Since the spallation rate is also $\nu_{\rm sp} \propto n_p$, the GRB dependencies largely cancel, and $E_{\rm Fe, crit}$ is always $O(10)$ GeV or higher\footnote{If $e^\pm$ pair production is important, cooling becomes more rapid and the critical energy becomes higher; see Section \ref{sec:shocks}.}. Equating $\nu_{\rm sp}$ to $ \nu_{Ae}$ and using the general relation $n_e = Z_i n_i$ yields the approximation
\begin{equation}
E_{\rm Fe, crit} \simeq \left[ \frac{2 \pi q^4 Z_A^2 Z_e^2 Z_i {\rm ln} \, \Lambda_{Ae} }{ \sigma_{\rm sp} m_e / m_A} \right]^{1/2} \approx 1.4 \times 10^{10} \, \rm{ eV},
\end{equation}
which is close to the value in Figure \ref{fig:timescales}. Note that this is a general critical energy derived from the competition between spallation and thermalization in a plasma consisting of electrons and ions of charge $Z_i$. It can also be applied to, e.g., the cocoon (Section \ref{sec:survival_cocoon}), where $Z_i >1$.

\begin{figure}[tb]
\includegraphics[width=3.25in]{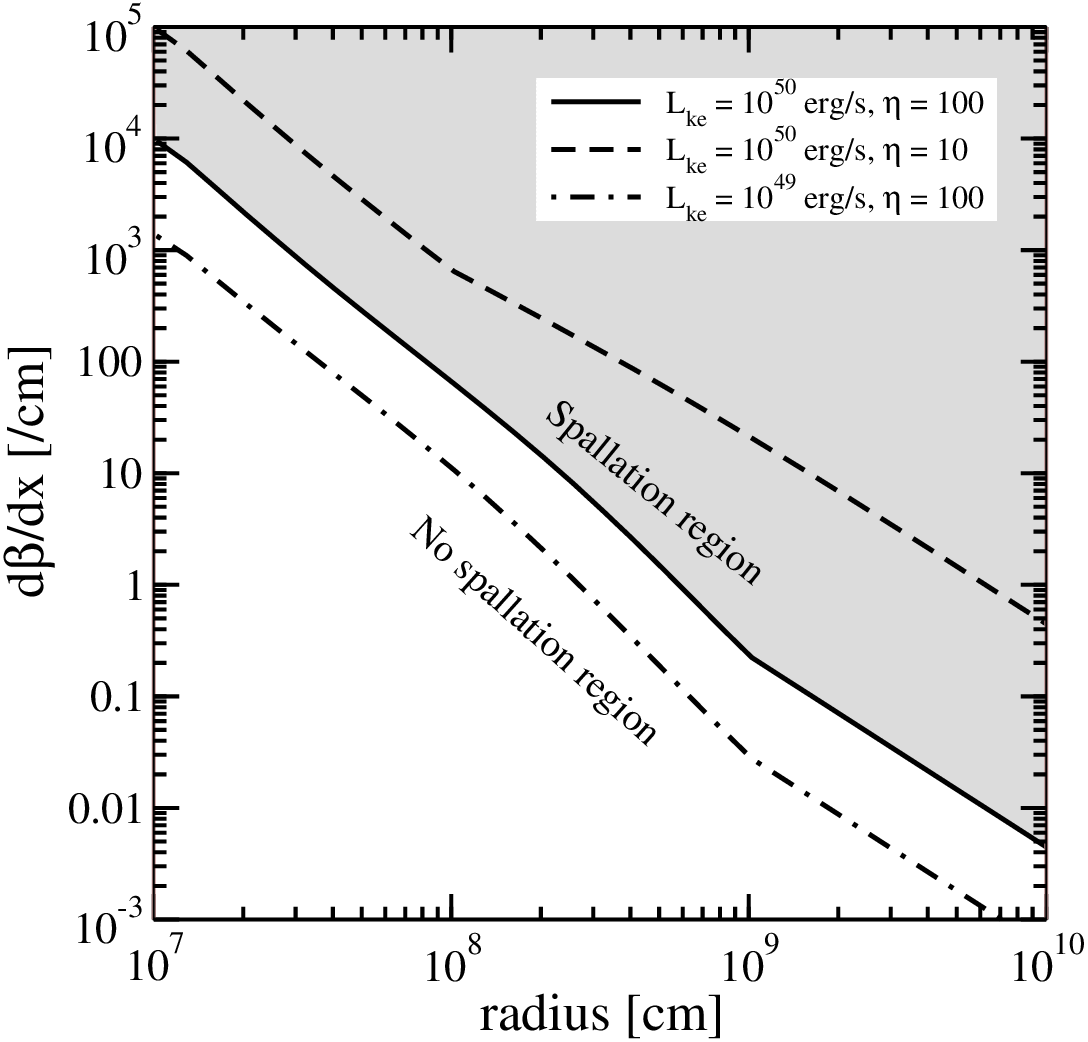}
\caption{\label{fig:shearconstraint} Upper limits on the $d\beta/dx$ that are required for Fe nuclei that move into the jet medium to survive spallation. Shown for various combinations of $L_{\rm ke}$ and $\eta$. For high $d\beta/dx$, the nuclei--jet proton collisions become more energetic and nuclei are destroyed. Shading shows this spallation region, drawn for our nominal $L_{\rm ke} = 10^{50} \, {\rm erg \, s^{-1}}$ and $\eta = 100$. \\}
\end{figure}

Let us denote the bulk velocity gradient at the cocoon--jet boundary by $d\beta/dx>0$, where $x$ is in the transverse direction with the origin at the boundary. As a nuclei moves into the jet in the $x$-direction, collisions with jet protons become increasingly more energetic, until spallation becomes energetically possible. However, the nuclei will also tend to thermalize with the jet plasma. Once thermalized, the collision between nuclei and jet ions will lack the energy to cause spallation. Thus, if spallation is slow enough compared to the thermalization rate, nuclei will survive. Since the spallation rate depends on the collision energy, an upper limit on $d\beta/dx$ can be derived. 

Consider a test nuclei moving in the $x$-direction with an initial velocity $v_{A,c} \sim 6 \times 10^7\, {T_{c,5}}^{-1/2} \, {\rm cm \, s^{-1}}$ corresponding to $T_c$, and let us work in the frame of the jet immediately surrounding the test nuclei. After moving a distance $\Delta x$, if the nuclei have not yet thermalized with the surrounding jet plasma, their velocity perpendicular to $x$ is Lorentz boosted by a frame change of $c (d\beta/dx) \Delta x$. As we showed in Figure \ref{fig:timescales}, nuclei survival requires the nuclei kinetic energy to be less than $E_{\rm Fe, crit}$. This yields
\begin{equation} \label{eq:shearlimit}
\frac{d\beta}{dx} \Delta x < \left[ 1- \left( \frac{m_A}{m_A + E_{\rm Fe, crit}}\right)^2 \right]^{1/2} \sim 0.7,
\end{equation}
where we have neglected the initial thermal kinetic energy of the nuclei as it is much smaller than $E_{\rm Fe, crit}$. Since $E_{\rm Fe, crit}$ depends weakly on the GRB parameters, Equation (\ref{eq:shearlimit}) does not show strong parameter dependencies either. Now, the distance $\Delta x$ is governed by thermalization. Since $\nu_{Ae}$ falls with energy, the thermalization distance grows with energy. Larger thermalization distances result in larger frame change boosts and place stronger constraints on $d\beta/dx$. So we consider the thermalization distance at $E_{\rm Fe, crit}$,
\begin{equation} \label{eq:RelaxationDistance}
\Delta x_{\rm therm} \simeq \frac{ v_{A,c}}{\nu_{Ae}(E_{\rm Fe, crit}) } \sim 6 \, T_{c,5}^{-1/2} \nu_{Ae,7}^{-1}  \, {\rm cm},
\end{equation}
where we have assumed that the energetic nuclei move in a straight line (in the $x$ direction) with velocity $v_{A,c}$. This is an upper limit, unless nuclei are carried by e.g., bulk flows faster than $v_{A,c} $. From Equations (\ref{eq:shearlimit}) and (\ref{eq:RelaxationDistance}) we can derive an upper limit on the $d\beta/dx$ required for nuclei survival. In other words, if the velocity gradient is steeper, the collision between nuclei and jet ions will become sufficiently energetic in the course of the nuclei moving a thermalization distance and spallation will occur. 

In Figure \ref{fig:shearconstraint}, we show the upper limits on $d\beta/dx$ for various GRB jet parameters all for a $\xi=2$ jet. The limits become more stringent with increasing radius, jet Lorentz factor, and inversely as the jet kinetic luminosity. These make sense, since these provide smaller target densities for nuclei to thermalize with and thus must yield stricter upper limits. For the same reason, the upper limits for a $\xi =1$ jet are relaxed compared to the $\xi=2$ jet. The breaks in Figure \ref{fig:shearconstraint} are due to the changing dependency of the jet bulk Lorentz factor on radius before and after saturation. 

We have made several simplifying assumptions that are conservative in nature. First, we have adopted a fixed particle density in the entire the cocoon--jet boundary equal to the jet particle density. In reality, the density in the cocoon--jet boundary will be larger due to a particle density gradient from the more dense non-relativistic cocoon to the more tenuous jet. By fixing to the jet density we have been conservative in estimating the upper limit on $d\beta/dx$. We also neglected $e^\pm$ pairs which at small radii will dominate over the $e^-$ associated with jet ions. As we detail in Section \ref{sec:shocks}, the production of $e^\pm$ pairs aids in nuclei survival because the pairs provide additional targets to which nuclei lose energy to, helping nuclei to cool on a shorter time scale. This would allow steeper velocity gradients. Finally, we have adopted the criterion that the spallation rate is smaller. This is perhaps conservative in light of the fact that a single spallation can still retain intermediate-mass nuclei composition. An alternate comparison we could have made would be to the spallation energy loss rate. Authors have adopted other criteria, e.g., \cite{MGH11} consider an optical depth $\lesssim 10$ to be acceptable. These relaxations will allow larger velocity gradients. 

We have also neglected magnetic fields at the jet--cocoon boundary. Magnetic fields remain highly uncertain but may be advected from the central engine or amplified in situ. For example, \cite{ZMW09} find that shear-driven instabilities induce macroscopic turbulence that leads to the amplification of magnetic fields which saturate at an energy fraction of $\epsilon_B \approx 10^{-3}$. If such processes also act efficiently inside the progenitor, the magnetic field can be as strong as $ (8 \pi \epsilon_B U_h )^{1/2} \sim 10^9$ G at $10^9$ cm. Strong magnetic fields can affect nuclei propagation in different ways depending on the field strength and power spectrum, as well as turbulence properties. In one limit, nuclei may propagate diffusively instead of rectilinearly. The effective transverse velocity of nuclei moving into the jet is then reduced, giving more time for thermal relaxation and allowing nuclei to survive larger velocity gradients. In the other limit, magnetic fields may trap nuclei. In this case, one may consider nuclei being carried by turbulence flows. Such flows can have a variety of velocities, lifetimes, and mixing efficiency. They may reach up to the sound speed $\sim c/\sqrt{3}$. From Figure \ref{fig:timescales}, nuclei moving at speed $c/ \sqrt{3}$ is in the spallation regime. However, it is marginal, and nuclei may still survive considering the conservative assumptions described above.

Finally, we have implicitly assumed that baryon entrainment occurs efficiently while the jet propagates through the star. We discuss the implications of the uncertain entrainment efficiency in the discussions section.

\subsubsection{Expected Velocity Gradient at the Jet--Cocoon Boundary}

The dynamics of relativistic GRB jets propagating through their progenitor stars have been extensively studied both analytically and numerically \citep{A00,ZM03,M03,ZWH04,MYNM06,MLB07}. While simulations confirm that the bulk velocity drops sharply at the jet boundary, with $\Gamma_j$ decreasing to order unity, a meaningful comparison with Figure \ref{fig:shearconstraint} is not yet possible. Even with state-of-the art numerical simulations, the desired transverse resolutions have not yet been reached. For example, in the simulations of relativistic jet propagation by \cite{ZM03}, the angular resolution is $0.25$ degrees, i.e., a few $10^7$ cm at radii of $\sim 10^9$ cm. Finer features are expected and seen with increasing simulation resolution \citep[e.g.,][]{MYNM06}, but simulations do not yet reach comparable scales as Figure \ref{fig:shearconstraint}.

However, it is clear that the velocity gradient should be connected to instabilities at the boundary. \cite{AIMU02} studied the stability properties of GRB jets propagating through a dense medium like the progenitor star. Through both numerical simulations and linear stability analysis, the authors find that the shear at the jet--cocoon boundary is responsible for instabilities that grow on very rapid time-scales. They show that the growth rate of non-homogeneous radial perturbations increases with the velocity gradient, i.e., $\nu_{\rm KH} \sim \Gamma_j |V'|$, where $|V'| $ is the initial velocity gradient. Therefore, a large $|V'| $ would cause the rapid growth of instabilities which would tend to smooth the velocity gradient to a smaller final velocity gradient $d\beta / dx$. Recall that nuclei survival requires the velocity gradient $d\beta/dx$ to be small over distance scales of $\sim \Delta x_{\rm therm}$. Thus, having a large initial gradient $|V'|$ may even help provide the necessary shallow $d\beta/dx$ for nuclei survival. For example, adopting the upper limit $\sim 0.2 \, {\rm cm^{-1}} $ (at $10^9$ cm; Figure \ref{fig:shearconstraint}) for $|V'|$, the instability growth rate $\nu_{\rm KH} \sim 6 \times 10^{11} \, {\rm s^{-1}}$ is much faster than the spallation and energy loss rates, so it is likely instabilities would grow and reduce the velocity gradient. For much smaller initial velocity gradients, instabilities may not have sufficient time to grow, but the small initial velocity gradient would anyways work positively for nuclei survival.

\section{Effects of bulk acceleration and dissipation} \label{sec:survival}

In the previous section, several candidates of the origin of jet nuclei were identified. However, even if nuclei are successfully loaded into the jet, they may be destroyed during the evolution of the jet because of bulk acceleration and/or dissipation. Here we discuss some of these effects.

\subsection{Neutrons during Bulk Acceleration} \label{sec:neutrons}

\begin{figure}[t]
\includegraphics[width=3.25in]{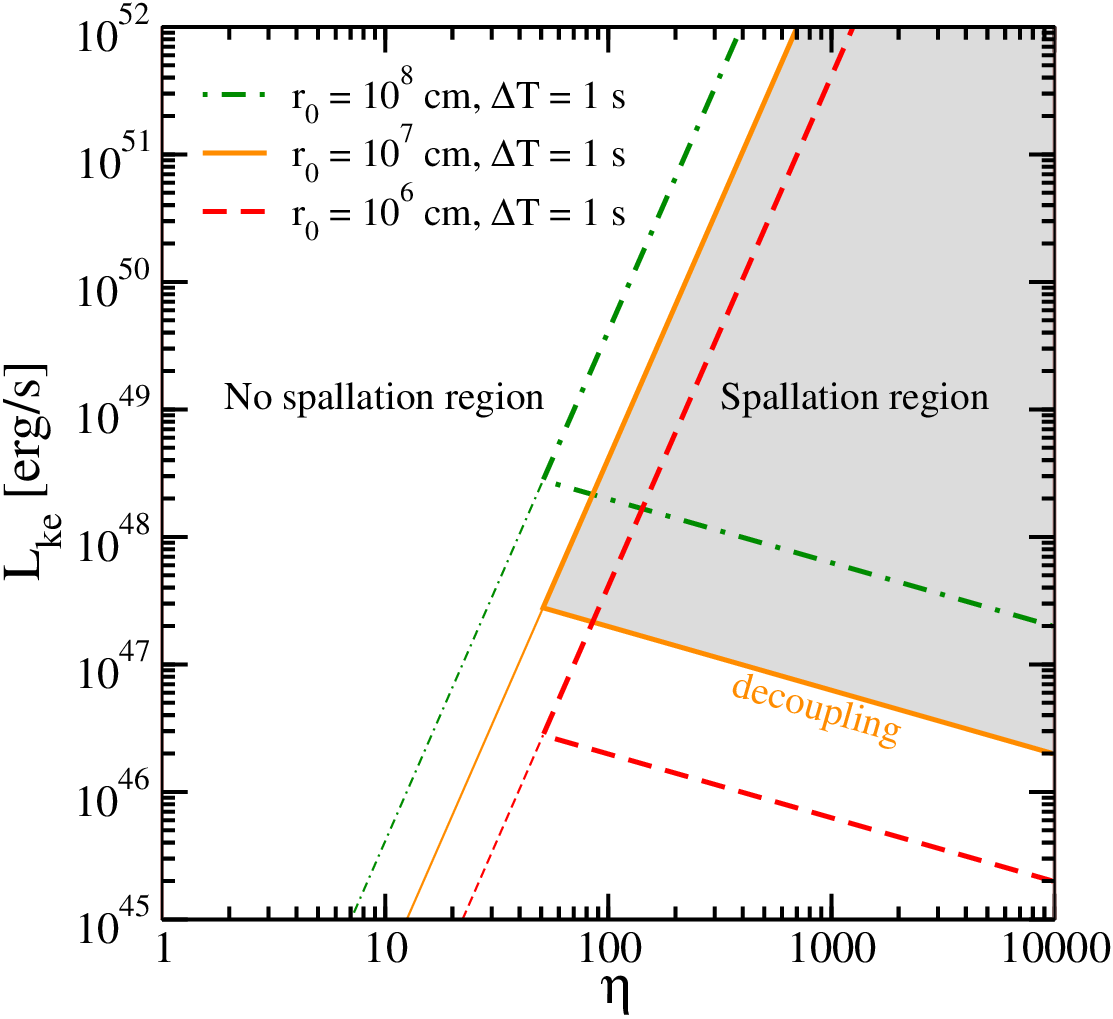}
\caption{\label{fig:neutrons} Constraints on jet parameters for Fe nuclei survival during jet bulk acceleration, including spallation caused by collisions with neutrons (lines increasing with $\eta$) and effects of neutron decoupling (lines decreasing with $\eta$, labeled), for several values of $r_0$. Shaded region results in nuclei spallation (shade shown for our nominal $r_0 = 10^7$ cm and $\Delta T=1$ s). \\}
\end{figure}

Initially, the neutron component of the jet is well coupled to the ion component by elastic collisions with a small relative velocity $\tilde{\beta} \ll 1$ that is insufficient to cause nuclei to break up.  However, the neutrons will lag behind the ions in the plasma during the bulk acceleration phase, if the collisions cannot keep up with the jet expansion. The velocity lag of neutrons relative to charged ions is $\Delta \Gamma / \Gamma \sim \tau_{\rm coll} /  \tau_{\rm exp}$, where $\Gamma_n = \Gamma - \Delta \Gamma$ is the neutron Lorentz factor, $\tau_{\rm coll} \simeq 1/(n_i \sigma_i \tilde{\beta} c)$ is the comoving collision time-scale, $\sigma_i \simeq \sigma_0 \bar{A}^{2/3} / \tilde{\beta}$ with $\sigma_0 \approx 3 \times 10^{-26} \, {\rm cm^2}$, and $\tau_{\rm exp} \simeq r / (\Gamma_j c)$ is the comoving expansion time-scale. The neutron--ion relative velocity is
\begin{equation} 
\tilde{\beta} = \frac{\beta - \beta_n}{1-\beta \beta_n} \sim \frac{\Gamma - \Gamma_n}{\Gamma} \sim \frac{\Gamma}{n_i \sigma_0 \bar{A}^{2/3} r} ,
\end{equation}
where $\beta_n$ is the neutron velocity. The relative velocity increases with radius during the bulk acceleration phase as $\tilde{\beta} \propto {L_{\rm ke}}^{-1} r^{2\xi-1}\eta $. Therefore, we require that at the end of the bulk acceleration phase ($r=r_{\rm sat}$), the relative velocity is smaller than $\beta_{\rm sp}$. This yields the condition for no spallation of
\begin{equation} \label{eq:etaspallation}
\eta < 220 L_{\rm ke,50}^{1/4} r_{0,7}^{-1/4},
\end{equation}
for $\xi = 2 $. To be conservative we have assumed an Fe composition for the jet: this leads to a larger $\tilde{\beta}$ than a proton jet, because of the longer collisional timescale. For a $\xi =1$ jet, the saturation radius is larger by a factor $\eta$ and the expansion time-scale is longer. Also, the higher jet density results in a faster collision time-scale. Both serve to relax the nuclei survival condition, yielding $\eta < 1300 {L_{\rm ke,50}}^{1/3}{r_{0,7}}^{-1/3}$. Similar conclusions hold for our adopted magnetically dominated jet model: the survival condition is $\eta < 480 {L_{\rm ke,50}}^{2/7}{r_{0,7}}^{-2/7}$. If the above conditions are not satisfied, the neutron density is typically high enough that nuclei are spalled very rapidly \citep{B03}. We plot the constraints for a $\xi = 2$ jet in Figure \ref{fig:neutrons}. 

An alternate way for nuclei survival is if neutrons decouple from the accelerating plasma. We define the decoupling radius as the jet radius when the separation between neutrons and ions grows larger than the radial width of the ejecta, $c \Delta T$, where $\Delta T$ is the jet pulse duration. In the lab frame, the separation is 
\begin{equation}
\int \tilde{\beta}(r) c \, dt =\int \frac{\tilde{\beta}}{ \beta_h} \, dr \propto L_{\rm ke}^{-3/2} r^{(5\xi-n)/2} \eta,
\end{equation}
(for $\xi \ne n/5$) which implies that decoupling is more likely to occur in low $L_{\rm ke}$ and high $\eta$ jets; for example, it does not occur for our nominal $L_{\rm ke} = 10^{50} {\rm erg \, s^{-1}}$ and $\eta=100$ jet. Let us therefore consider a jet of luminosity $L_{\rm ke} = 10^{47} \, {\rm erg \, s^{-1}}$ and terminal Lorentz factor $\eta= 100$. Note that from Equation (\ref{eq:etaspallation}), it can been seen that spallation will occur in such a jet. Now, the spallation radius (defined as when $\tilde{\beta} = \beta_{\rm sp}$) and the decoupling radius are, respectively,
\begin{eqnarray}
&&r_{\rm sp} \sim 2.9 \times 10^8 L_{\rm ke,47}^{1/3} \eta_2^{-1/3} r_{0,7}^{2/3} \, {\rm cm} \\
&&r_{\rm de} \sim 2.7 \times 10^8 L_{\rm ke,47}^{3/7} \eta_2^{-2/7} r_{0,7}^{4/7} \left( \frac{\Delta T}{1 \,{\rm s}} \right)^{2/7} \, {\rm cm},
\end{eqnarray}
both for $\xi =2$. Requiring that decoupling precedes spallation yields
\begin{equation} \label{eq:etadecoupling}
\eta < 400 L_{\rm ke,47}^{-2} r_{0,7}^{2} \left( \frac{\Delta T}{1 \,{\rm s}} \right)^{-6}.
\end{equation}
As expected, decoupling becomes important for sub-luminous jets. For example, a small $L_{\rm ke}$ strongly limits the range of $\eta$ for which spallation does not occur [Eq.~(\ref{eq:etaspallation})], but the consideration of neutron decoupling allows a much wider range to be acceptable [Equation (\ref{eq:etadecoupling})]. However, we caution the strong dependencies of Equation (\ref{eq:etadecoupling}) on GRB parameters. We conclude that decoupling may play a role but its importance is strongly parameter dependent. Note that for a $\xi =1$ jet the particle density is larger and the relative velocity $\tilde{\beta}$ is significantly slower, so that decoupling typically does not occur while the jet is propagating through the stellar core. In Figure \ref{fig:neutrons}, we show constraints on $L_{\rm ke}$ and $\eta$ for several values of $r_0$, for a $\xi =2 $ jet with $\Delta T = 1$ s.

In conclusion, nuclei can survive over a significant range of $L_{\rm ke}$ and $\eta$ because of slow neutron--ion relative velocities. For low $L_{\rm ke}$ (and $\Delta T$), decoupling may occur before spallation so that a wider range of $\eta$ is allowed. For canonical values of $L_{\rm ke}$ and $\eta$, collisions just reach spallation thresholds. For more collimated jets, a larger GRB parameter range is allowed.

\subsection{Dissipation and Particle Heating at Oblique Shocks} \label{sec:shocks}

Until now we have parameterized the jet cross section by Equation (\ref{eq:crosssection}) which can be adapted to any morphology, e.g., conical (constant jet opening angle) or funnel (jet opening angle $\propto r^{-1/2}$) behavior. However, recent numerical and analytic works of jet propagation show that the jet morphology is not a free parameter but correlated with the jet's interactions with the cocoon. According to the analytic work of \cite{BNPS11}, which is based on the jet--cocoon solutions of \cite{BC89} and \cite{M03}, the jet morphology is determined by the jet energy density, its opening angle, and the density of the external medium. For a jet propagating through a dense material such as the progenitor star, the cocoon pressure is sufficiently strong and compresses the jet. Oblique collimation shocks form which generate the pressure needed to counterbalance the cocoon's pressure. The shocks can form early in jet propagation but converge as the entire jet is shocked at a radius of approximately $r_{\rm sh} \sim 0.1 r_h$, where $r_h$ is the jet head radius \citep[e.g., Figure 2 of][]{BNPS11}. 

For a sufficiently fast jet ($\Gamma_j > 1/\theta_j $), the Lorentz factor of the jet after being shocked is $\Gamma_s \sim 1/ \theta_j$ \citep{BNPS11}, which for our GRB parameters $\theta_j \approx 0.13 {\Omega_{0,-1}}^{1/2}$ and thus $\Gamma_s \sim 8 \, { \Omega_{0,-1}}^{-1/2}$. The jet ions passing the shock may obtain high random Lorentz factors corresponding to the relative Lorentz factor $\sim \Gamma_j / (2 \Gamma_s) $ in a similar way to the reverse shock (Section \ref{sec:survival_cocoon}). Such ions would collide with other jet ions and any nuclei could be spalled. Here, we consider the spallation rate and energy loss rate to see whether jet nuclei are actually destroyed. To place conservative limits, we consider the fate of Fe ions within a jet plasma consisting of mainly protons. This assumption yields the highest spallation target density; the spallation rate is then $\nu_{\rm sp} \simeq n_p \sigma_{\rm sp} c$, yielding ${\nu_{\rm sp}}^{(2)}\sim 3 \times 10^{7} {L_{\rm ke,50}} {r_{\rm sh,9}}^{-3} {\eta_2}^{-1} \, {\rm s^{-1}}$ ($\xi =2$) or ${\nu_{\rm sp}}^{(1)}\sim 3 \times 10^{10} {L_{\rm ke,50}} {r_{\rm sh,9}}^{-3/2} {\eta_2}^{-1} \, {\rm s^{-1}}$ ($\xi =1$).

To estimate the Fe-electron collisional cooling rate, we consider in addition to the electrons associated with jet protons the production of $e^\pm$ pairs. This directly increases the Fe cooling rate since $\nu_{Ae} \propto n_e$. In fact, it is important for the survival of relativistic nuclei, since we showed in Section \ref{sec:entrainment} that only mildly relativistic Fe ions can survive when electrons associated with jet protons are considered. For a radiation temperature $T_r$ the equilibrium pair density is \citep{SP90}
\begin{equation}
n_\pm \approx 4.4 \times 10^{30} (T_r / m_e)^{3/2} e^{-m_e / T_r}.
\end{equation}
We estimate the pair density in the jet by substituting the jet radiation temperature. For example, for a $\xi=2$ jet in the bulk acceleration phase, ${T_j}^{(2)} \sim 180 {L_{\rm ke,50}}^{1/4} {r_{9}}^{-1}$ keV and the resulting pair density exceeds those associated with jet protons by a factor of approximately $10^4$ at $10^8$ cm. However, by $\sim 10^{8.5}$ cm the $T_j $ has fallen too low and pairs make a negligible contribution to the total electron density. For a narrower jet, the jet temperature remains high longer so that pairs make meaningful contributions out to larger radii, e.g., for $\xi = 1$ and canonical jet parameters, out to $\sim 10^{10}$ cm. Note that the jet radiation temperature can be high but photodisintegration is still slower than the spallation rate. 

Let us first discuss when the fast jet approximation of \cite{BNPS11} is valid, i.e., $r_{\rm sh} $ is larger than $8 \times 10^7$ cm ($\xi = 2$) or $6.4 \times 10^8$ cm ($\xi = 1$), both for our canonical choices of $r_0$ and $\Omega_0$. Whether collisional cooling prevents spallation depends quantitatively on the Lorentz factor of the Fe ions and the electron density. For example, at $r_{\rm sh} = 10^8$ cm, the Fe ion cooling rate is ${\nu_{Ae}}^{(2)} \sim 1 \times 10^{12}  {E_{\rm Fe,12}}^{-1} \, {\rm s^{-1}}$ for our canonical GRB parameters and $\xi = 2$. Here we take the electron temperature to be equivalent to the radiation temperature, because electrons reach local thermodynamic equilibrium on much shorter time scales compared to the ion--electron energy loss time scale. Comparing ${\nu_{Ae}}^{(2)}$ to the spallation rate ${\nu_{\rm sp}}^{(2)}$, we see that Fe ions will undergo spallation if their energies are above $E_{\rm Fe} \sim 4 \times 10^{13} \, {\rm eV}$, or a Lorentz factor of $\sim 800$. Thus for our canonical parameters, Fe ions easily survive at $r_{\rm sh} = 10^8$ cm. 

\begin{figure}[t]
\includegraphics[width=3.25in]{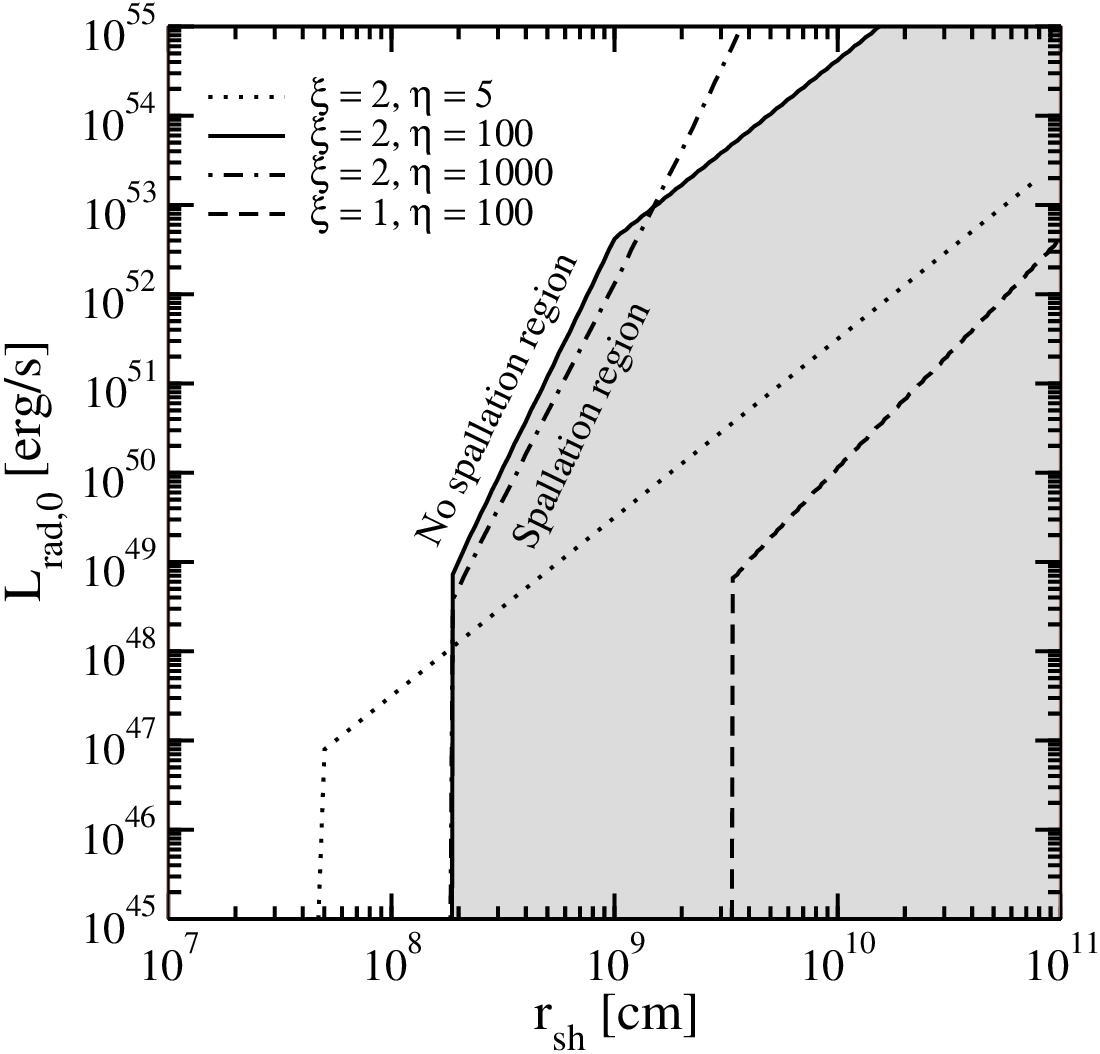}
\caption{\label{fig:shocksurvival} Constraints on jet luminosity for Fe nuclei survival in collimation shocks. Shaded region results in nuclei spallation (shown for a $\xi = 2$ and $\eta = 100$ jet). A narrower jet is more favorable for nuclei survival due to its higher temperature (dashed line). The $\eta = 5$ jet (dotted) is simply an illustration where the post-shocked jet Lorentz factor is deliberately fixed so that nuclei are spalled; it is possible that the post-shocked jet is faster in which case nuclei will not be spalled. For all other jets, the post-shock dynamics are calculated according to the collimation shock model of \cite{BNPS11}, see the text.}
\end{figure}

However, as $r_{\rm sh}$ increases, the pair density drops and survival becomes increasingly difficult. In Figure \ref{fig:shocksurvival}, we show the jet luminosity required for nuclei survival as a function of $r_{\rm sh}$. This is determined by calculating the Fe Lorentz factor as a function of $r_{\rm sh}$, and then the necessary pair density so that the energy loss rate $\nu_{Ae}$ is faster than the spallation rate $\nu_{\rm sp}$. It is clear that higher luminosities are required for survival at larger radii, because the temperatures are lower. For a similar reason, a narrower jet is more favorable. A larger value of $\eta$ is only slightly more favorable mainly due to the smaller spallation target density it implies. Note that the importance of pairs can be noticed: for $L_{\rm rad} = 10^{50} {\rm erg \, s^{-1}}$, nuclei survive out to $r_{\rm sh} \approx 10^{8.5}$ cm ($\xi =2$) or $r_{\rm sh} \approx 10^{10}$ cm ($\xi =1$); these radii are the radii where pairs cease to make a contribution. At the small radius end, the required luminosity suddenly drops because the Fe Lorentz factor tends to unity as the radius where the fast jet approximation becomes invalid is reached. 

The model of \cite{BNPS11} cannot be applied for radii where the fast jet approximation is not valid, i.e., $r_{\rm sh} < 8 \times 10^7$ cm ($\xi = 2$) or $r_{\rm sh} < 6.4 \times 10^8$ cm ($\xi = 1$), for our parameters. However, that is not to say oblique shocks will not occur in this region. The stellar pressure is typically much greater than the jet pressure so that the jet is pressure confined by the star, a setup conducive to oblique shock formation. In fact, much of the early jet phase can be shocked and supply the cocoon material. We do not attempt to model the dynamics of these jets. However, we can reason that due to the high jet radiation temperatures, it is unlikely nuclei are destroyed. In these radius ranges, the pair density exceeds the proton density by $\sim 10^4$ and cooling is rapid. For example, at $r_{\rm sh}=10^{7.5}$ cm ($\xi = 2$) or $r_{\rm sh}=10^8$ cm ($\xi = 1$), the cooling rates are ${\nu_{Ae}} \sim 5 \times 10^{13}  {E_{\rm Fe,12}}^{-1} \, {\rm s^{-1}}$, and spallation only occur for highly relativistic nuclei with Lorentz factors greater than $\sim 1000$. 

Similarly, jets with $\eta < 1 / \theta_j$ do not satisfy the fast jet approximation, but oblique shocks are nonetheless expected as the jet expands into the stellar material and later on into the cocoon material. In this case we treat $\Gamma_s$ as a free parameter. If $\Gamma_s \sim \eta$, then nuclei are not spalled because the relative Lorentz factor between the jet and post-shocked jet is small. On the other hand, if $\Gamma_s \ll \eta$, the nuclei can attain mildly relativistic energies and can be spalled. For example, nuclei will survive in a $\eta = 5$ jet if $\Gamma_s \approx 2.3$ or higher. In Figure \ref{fig:shocksurvival} we shown as an example the case where $\Gamma_s $ is fixed such that nuclei are spalled (shown for $\Gamma_s = 2$; dotted). As expected, the smaller relative Lorentz factor results in a weaker requirement than our canonical ($\eta = 100$) jet of the same morphology (dashed). 

Oblique shocks can also occur after jet collimation as the jet propagates through the progenitor, but the survival of nuclei depends largely on the post-shocked jet Lorentz factor. If $\Gamma_s \sim \eta$ as is usually expected, nuclei are likely not destroyed. 

Note that in all the above estimates we have conservatively assumed a proton-dominated jet and considered the survival of a (minor) Fe component. If the jet is mostly Fe dominated, the spallation target density decreases by $A=56$ and improves Fe survival prospects. Also, we have conservatively considered the spallation rate. If we relax this and consider instead the spallation energy loss rate, i.e., we allow more than one spallation event, the limits for nuclei survival will be relaxed. 

To conclude, jet nuclei may be spalled if jet ions obtain relativistic random velocities at oblique collimation shocks. However, when the shock radius $r_{\rm sh}$ is small, the pair density in the jet can be sufficiently high that nuclei energy loss is faster than spallation. For our canonical jet parameters, this condition is realized for shock radii less than $\sim 10^{10}$ cm for an initially narrow jet (which may be realized if the jet is initially confined by the rotational funnel geometry of the progenitor) or less than $\sim 10^{8.5}$ cm for an initially conical jet. Nuclei that are entrained at larger radii (i.e., $r> r_{\rm sh}$) may not affected by the oblique collimation shocks and are not spalled. Jets with small $\eta$ imply small Fe ion Lorentz factors and are thus more favorable for survival. 

\subsection{Dissipation and Particle Acceleration at Emission Regions} \label{sec:dissipation}

GRB emissions typically consist of prompt gamma-ray emission and afterglow emission. The former is attributed to non-thermal or quasi-thermal emissions produced via some internal dissipation, e.g., internal shocks or magnetic reconnections. In the standard optically thin synchrotron scenario of the internal shock model, inhomogeneities in the jet cause internal shocks within the jet, where the relative bulk kinetic energy is dissipated and accelerated electrons radiate gamma rays. In this scenario, it is natural to expect that ions are accelerated as well. The maximum energy to which ions are accelerated is roughly determined by requiring that the acceleration timescale is shorter than the dynamical and any energy loss timescales. Previous works have shown that the acceleration of nuclei to UHECR energies is possible for certain ranges of GRB parameters when internal dissipation happens in the optically thin regime \citep[][]{MINN08,WRM08}.

The conditions required for the accelerated nuclei to survive in the prompt GRB photon field are much more stringent than those for acceleration.  The comoving photodisintegration time scale for a nucleus moving through an isotropic photon background is~\citep[e.g.,][]{MINN08,WRM08}
\begin{equation}
t_{A\gamma}^{-1} = \frac{c}{2 \gamma_A^2} \int^\infty_{\bar{\varepsilon}_{\rm th}} d \bar{\varepsilon} \,\, \bar{\varepsilon} \sigma_{A\gamma}(\bar{\varepsilon}) \int^\infty_{\bar{\varepsilon}/2\gamma_A} d\varepsilon \,\,  \frac{1}{\varepsilon^2} \frac{dn}{d\varepsilon},
\end{equation}
where quantities such as $\bar{\varepsilon}$ are defined in the nucleus rest frame, $\gamma_A$ is the nucleus Lorentz factor, $\bar{\varepsilon}_{\rm th} \approx 7.6$ MeV is the threshold photon energy for photodisintegration of an iron nucleus, and $dn/d\varepsilon$ is the differential photon spectrum.  Adopting a comoving broken power law for the prompt GRB photon spectrum,
\begin{equation} \label{eq:GRBprompt}
\frac{dn}{d\varepsilon} \approx \frac{\epsilon_e L_{\rm ke} e^{-\varepsilon/\varepsilon_{\rm max}}}{ 5 \Sigma \eta^2 c \varepsilon_b^2} \left\{ 
\begin{array}{ll}
(\varepsilon/\varepsilon_b)^{-1} & \quad  \varepsilon_{\rm min} < \varepsilon < \varepsilon_b \\
(\varepsilon/\varepsilon_b)^{-2.2} & \quad  \varepsilon_b < \varepsilon < \varepsilon_{\rm max},
\end{array}\right.
\end{equation}
where $\varepsilon_b = \varepsilon_{\rm b,obs} / \eta$ is the break energy, the factor $5$ in the denominator arises because the luminosity at the break energy is approximately $1/5$ the total luminosity, $\varepsilon_{\rm b,obs} \approx 1$ MeV, $\varepsilon_{\rm min} = 1$ eV, and $\varepsilon_{\rm max} = 10$ MeV, the comoving photodisintegration timescale is estimated to be
\begin{equation}
t_{A\gamma} \sim 300 \epsilon_{e,-1}^{-1} L_{\rm ke}^{-1} r_{15}^2 \eta_2\varepsilon_{\rm b,obs,6} \,{\rm s},
\end{equation}
where we have adopted $\xi = 2$ since the jet becomes conical after jet break out. We require that $\tau_{\rm exp} < t_{A\gamma}$ for UHECR nuclei to survive photodisintegration in the system, which yields the constraint
\begin{equation}
\eta > 110 \epsilon_{e,-1}^{1/2} L_{\rm ke,50}^{1/2} r_{15}^{-1/2} \varepsilon_{\rm b,obs,6}^{-1/2}.
\end{equation}

\begin{figure}[tb]
\includegraphics[width=3.25in]{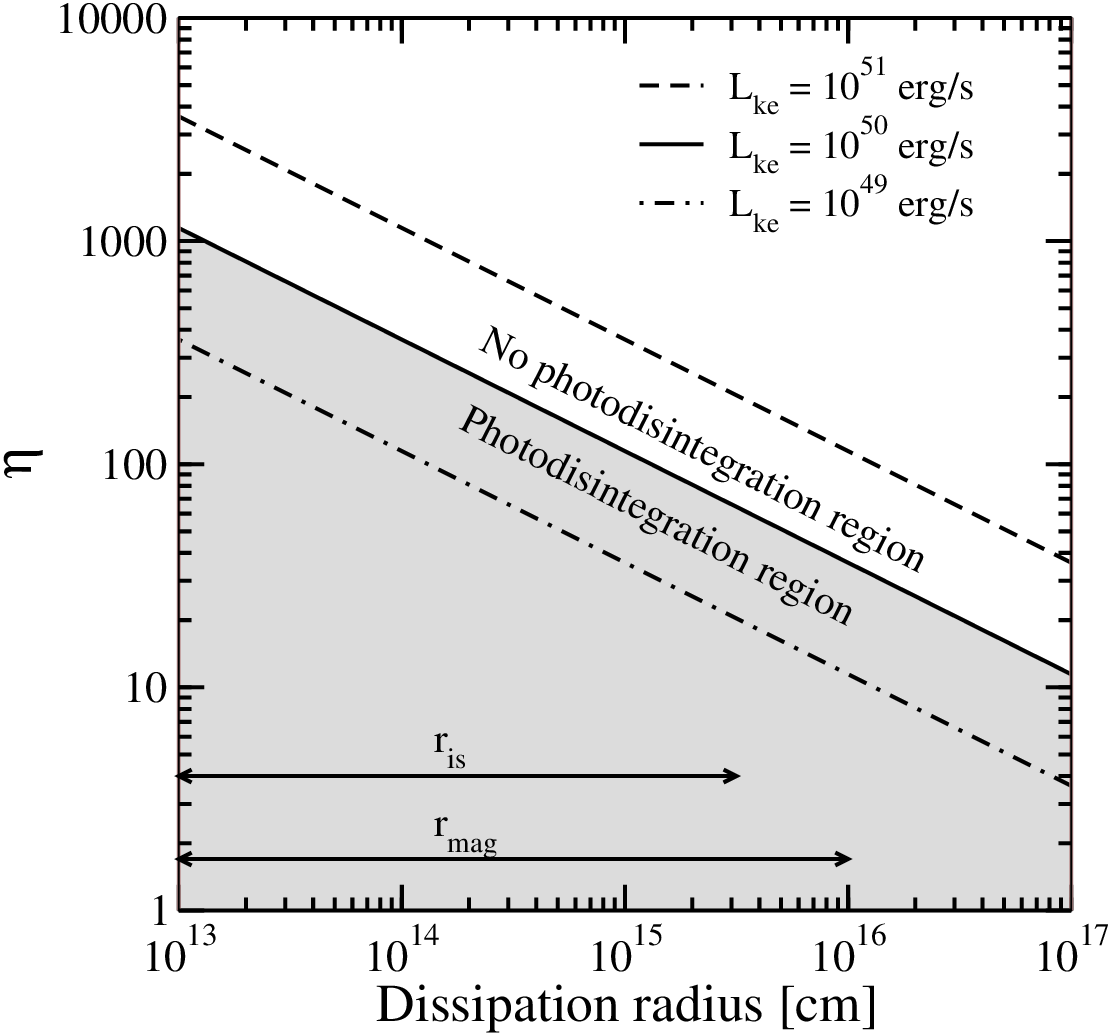}
\caption{\label{fig:Dissipation} Constraints on jet $\eta$ and internal dissipation radius for Fe survival at the dissipation region. GRB jets of smaller luminosity and larger $\eta$ allow smaller internal dissipation radii to satisfy nuclei survival conditions. For illustration, we indicate typical dissipation radii for the internal shock scenario ($r_{\rm is}$) and the magnetic reconnection scenario ($r_{\rm mag}$) by arrows. \\}
\end{figure}

We plot this in Figure \ref{fig:Dissipation} as a function of the dissipation radius, i.e., emission radius. We emphasize that the optical depth for photodisintegration highly depends on the dissipation radius that is estimated to be $r_{\rm is} \approx 2c \, \delta t \, \eta^2$ for the internal shock case. Typical internal shock radii are $\sim {10}^{13}$--${10}^{15.5}$~cm \citep[e.g.,][]{NP02}, so survival of UHECR nuclei is possible only for relatively large dissipation radii, although this is relaxed for sub-luminous GRBs~\citep{MINN08}. Typical radii for magnetic dissipation scenarios similarly span a wide range but can reach up to $\sim 10^{16}$ cm \citep[e.g.,][]{D02,GKS11,M11}. In Figure \ref{fig:Dissipation} we show these ranges of dissipation radii for reference, noting that there are substantial uncertainties. Note that in the photospheric emission scenario, the emission radius is typically expected to be much smaller, where high-energy nuclei break up via both spallation and photodisintegration and instead high-energy neutrinos are generated~\citep{M08}.   

Dissipation and emission also occur later on, when the jet expands into the interstellar medium and sweeps up sufficient baryons to enter the Blandford--McKee phase.  The forward and reverse shocks are collectively referred to as external shocks, where ions and electrons can be accelerated.  The observed non-thermal photons are attributed to synchrotron emission from relativistic electrons accelerated at such shocks.  Compared to the case of the prompt emission, nuclei interact with softer photons, and it has been shown that the survival of UHECR nuclei is also possible \citep[][]{MINN08,WRM08}.

\section{Discussions and Conclusions} \label{sec:conclusion}

In this work, we focus on the fate of nuclei in a relativistic jet accompanying a CCSN. We consider three sources of jet nuclei, (1) loading at the jet base during jet launch, (2) in situ explosive nucleosynthesis, and (3) entrainment of external nuclei during jet propagation. We first discuss the conditions for nuclei survival in each of them. For initial loading, we find that nuclei survive if the jet launch radius is greater than $r_0 \sim 10^8$ cm or the initial radiation luminosity is less than $L_{\rm rad,0} \sim 10^{48} \, {\rm erg \, s^{-1}}$ (Figure \ref{fig:result}). These conditions may be satisfied by magnetic models of classic GRBs or models of sub-luminous GRBs. If nuclei are destroyed at jet launch into free nucleons, the material is fused into nuclei, but the freezeout abundance contains significant heavy nuclei only for low entropy jets \citep{B03,I03,MGH11}. Finally, whether external nuclei survive during entrainment into the jet depends critically on the velocity gradient at the cocoon--jet boundary (Figure \ref{fig:shearconstraint}). If the gradient is too steep, collisions between external nuclei and jet ions become too energetic and external nuclei are spalled. On the other hand, if the gradient is shallow enough, the nuclei thermalize with the jet plasma before spallation and survive. Since the growth rate of shear-driven instabilities at the cocoon--jet boundary is larger than the nuclei energy loss or spallation times, the velocity gradient is expected to be shallow enough to allow nuclei to survive. 

Next we investigate the conditions for nuclei survival once loaded into the jet. First, collisions with neutrons become energetic during the jet bulk acceleration phase and can cause nuclei spallation if the GRB terminal Lorentz factor is larger than $\eta \sim 220 {L_{\rm ke,50}}^{1/4} {r_{0,7}}^{-1/4}$ (for a conical jet) or $\eta \sim 1300 {L_{\rm ke,50}}^{1/3} {r_{0,7}}^{-1/3}$ (for a funnel jet); for smaller $\eta$, spallation is energetically prohibited. Also, we identify the parameter space where the neutrons decouple before causing spallation. We find that it is strongly parameter dependent but tends to help with small $L_{\rm ke}$ jets (Figure \ref{fig:neutrons}). Second, we show that nuclei can survive at oblique shocks if $e^\pm$ pairs are produced such that the nuclei collisional energy loss rate is competitively high. Such conditions are typically realized for shock radii $r_{\rm sh}$ less than $ \sim 10^8$--$10^{10}$ cm (Figure \ref{fig:shocksurvival}). A smaller $\eta$ works positively for nuclei survival, as does a post-shocked jet Lorentz factor $\Gamma_s$ that is close to the pre-shocked jet Lorentz factor. 

Based on the above results, we can consider the sources and survival of nuclei in three distinct sets of GRB models. These are summarized below and in Table \ref{table:results}.

First, we conclude that nuclei from initial loading and explosive nucleosynthesis are disfavored in the HL GRB fireball scenario \citep[see also][]{B03}; instead, nuclei may come from entrainment during jet propagation. Once loaded, nuclei may survive collisions with neutrons. Survival in oblique collimation shocks is possible while $r_{\rm sh}$ is less than $\sim 10^8$--$10^{10}$ cm. After $r.p.m._{\rm sh}$ reaches these values, nuclei entrained at smaller radii ($r < r_{\rm sh}$) may be spalled as they cross the shocks, while nuclei that are entrained at larger radii ($r > r_{\rm sh}$) may survive. Successfully entrained nuclei can survive the emission region provided the dissipation radius is large enough (Figure \ref{fig:Dissipation}; see also \citealt{MINN08,WRM08}).

On the other hand, multiple sources of nuclei are possible for magnetic-dominated scenarios of HL GRBs. For example, nuclei may survive at initial loading, because for a given GRB luminosity the initial radiation energy can be lower. Even if nuclei are spalled at loading, the subsequent explosive nucleosynthesis can lead to Fe-group elements and possibly beyond \citep[][]{MGH11}. Entrainment can work too, if as in recent models the jet is collimated by a cocoon. The slower bulk acceleration compared to fireballs means collisions with neutrons are generally less destructive, and oblique shock are expected to be weaker, although definite statements are dependent on the specific model of magnetic dissipation. 

Similarly, there are multiple potential nuclei sources for sub-luminous GRBs and jet-driven CCSN. Their smaller luminosities and bulk Lorentz factors work positively for nuclei survival at initial loading and for explosive nucleosynthesis \citep{I03}. Although the low $L_{\rm ke}$ works negatively for spallation by neutrons, this is largely compensated by the lower $\eta$; as a result, nuclei are typically not destroyed (for the smallest $L_{\rm ke}$, neutron decoupling can also be effective). The low $\eta$ also works positively for nuclei survival in oblique collimation shocks, because the nuclei Lorentz factors are small. However, note that nuclei can still be spalled at large $r_{\rm sh}$ if $\Gamma_s \ll \eta$. Finally, though uncertainty is large, survival at dissipation is easier than in HL GRBs (Figure \ref{fig:Dissipation}; see also \citealt{MINN08,WRM08}). 

\begin{deluxetable}{lccc}
\tablecaption{Summary of Nuclei Survivability \label{table:results}}
\tablehead{ 							&	HL GRB 	& 	HL GRB 	&	LL GRB	\\
									&	Fireball	&	Magnetic	&				}
\startdata
Source: initial loading					&	N		&	Y		&	Y		\\
Source:  nucleosynthesis					&	N		&	Y		&	maybe	\\
Source:  entrainment						&	maybe	&	maybe	&	maybe	\\
\tableline
$n$ collisions					&	Y		&	Y		&	Y		\\
Collimation shocks\tablenotemark{a}	&	maybe	&	maybe	&	maybe	\\
Emission region\tablenotemark{b}	&	Y		&	Y		&	Y		
\tablenotetext{a}{For all cases, other oblique shocks may lead to nuclei destruction depending on shock velocity.}
\tablenotetext{b}{For emission radius $10^{15}$ cm.}
\tablecomments{``N'' denotes not possible for typical parameters; ``maybe'' possible depending on model, and ``Y'' possible for typical parameters. Typical parameters are taken to be $L_{\rm ke} = 10^{50} \, {\rm erg \, s^{-1}}$, $\eta = 300$, $\xi =2$, and $r_0 = 10^7 \, {\rm cm}$ for HL-GRBs and $L_{\rm ke} = 10^{47} \, {\rm erg \, s^{-1}}$, $\eta = 10$, $\xi =2$, and $r_0 = 10^7 \, {\rm cm}$ for LL GRBs. For entrainment, we consider the currently unconstrained jet--cocoon velocity gradient to be model-dependent.}
\end{deluxetable}

As seen in this work, jets accompanying CCSN may contain heavy nuclei that originate from the stellar core, disk wind, and/or low entropy (kinetically or magnetically dominated) jets. A high abundance of nuclei is attractive in view of recent reports of a heavy-ion dominated composition of the highest-energy UHECR. If AGNs are the sources of UHECRs, rigidity-dependent acceleration models naturally predict a heavy-ion composition at the highest energies, but this seems inconsistent with the null observation of an excess of protons at lower energies \citep{Auger11}. On the other hand, nuclei-rich UHECRs could be realized in GRBs and CCSN, as long as the accelerated particles are injected without being broken up. 

However, GRBs and relativistic CCSN scenarios may also have several issues such as an energy crisis and cosmic-ray escape problems. That an energy-crisis problem may be serious for classical HL GRBs has been claimed in view of \textit{Fermi} observations \citep[e.g.,][]{EGP10}, although the universality of this claim depends on the uncertain local rate of classical GRBs, the efficiency of cosmic ray acceleration, and the slope of the injected UHECR spectrum \citep[e.g.,][]{LD07,MINN08,W10}. It is interesting to note that magnetic GRB models may predict flatter accelerated cosmic ray spectra which could help to avoid an energy crisis \citep[e.g.,][]{RL92,M11}, although it remains to be seen whether it can also offset the issue of smaller baryonic loading in magnetic jets. Furthermore, sub-luminous GRBs may contribute to the observed UHECR flux~\citep[][]{MINN06}. The expected rate of CCSN with relativistic ejecta (whether it is jet-like or not) seems comparable to that of sub-luminous GRBs~\citep[][]{C11}, and both jet and relativistic CCSN (which may be jet-driven) can provide UHECR nuclei~\citep[e.g.,][]{MINN08,WRM08}.

We emphasize that we have attempted to place conservative constraints on nuclei survival. This means we have not always used the same jet composition in each investigation. For example, when we consider neutron collisions we adopt a nuclei-dominated jet so that the relative velocity is highest and spallation occurs most readily; and when we consider oblique shocks we adopt a proton-dominated jet so that the spallation rate is maximal. In the same spirit, we assume that a single spallation effect must be avoided, when in fact a few spallations may still maintain a heavy or intermediate nuclei composition depending on the initial composition. Relaxing our criterion and allowing, e.g., the spallation optical depth to be a few, relaxes the nuclei survival constraints described above. This would naturally lead to a final composition of more intermediate nuclei.

We have been simplistic on several issues. We have only considered two kinds of jets, and also not considered progenitor dependencies. But the most important is our implicit assumption that loading and entrainment of external baryons occur efficiently. Jets accompanying CCSN can have a wide range of baryonic content, with GRB jets among the cleanest with fewest baryons. GRB jet models achieve this in different ways. For example, magnetic-dominated jets that are powered by energy release on field lines where baryons are confined can plausibly contain very few initial baryons \citep[e.g.,][]{LB03}. Thus, although nuclei are more likely to be present at the launch sites of magnetic-dominated jets (Table \ref{table:results}), they may not be efficiently loaded. However, we stress that such a statement remains highly model dependent. Later entrainment is attractive in the sense that jets \emph{must} load baryons at some point, and propagation through the baryon-rich progenitor is an unavoidable course of jet evolution. The entrainment efficiency is expected to depend on the instability mechanism, thermodynamic parameters of the cocoon and jet, magnetic field, and so on, and thus evolve. To make quantitative estimates of these and other effects requires a detailed model of entrainment and is beyond the scope of the present paper. In this work, we investigated the process by which nuclei may survive during loading. Although the specific GRB model may change in the future, the physical processes by which nuclei survive should still hold.

We can however speculate about the composition assuming entrainment is the main process of jet baryon loading. In an oversimplified picture where entrainment occurs equally efficiently during the entire propagation of the jet, the jet composition will be similar to the total matter swept up in propagation. However, this is likely strongly modified. First, explosive nucleosynthesis is expected to increase the nuclei abundance of the swept up material. Indeed, high $^{56}$Ni masses have been observationally supported by nearby highly luminous CCSN \citep{I98,WES99,WH03}, and observational hints of jet-mixed nuclei have been observed \citep[e.g.,][]{S12}. Second, nuclei tend to be more easily destroyed during the later phases of jet evolution, in particular as the jet head velocity exceeds $\beta_{\rm sp }\sim 0.14 $, and/or when the oblique shock radius exceeds $r_{\rm sh} \sim 10^{10}$ cm. Nuclei that lie at radii $\gtrsim 10^{10}$ cm, mainly intermediate nuclei such as carbon, are thus more likely to be spalled before entering the cocoon.
 
Finally, it is relevant to identify observational signatures of heavy nuclei. For protons, neutrinos are produced when accelerated protons interact with photons and/or target protons and produce charged pions. The neutrino signatures have been well studied in various contexts. If collisionless shocks can be formed inside the progenitor star, GeV--TeV neutrinos might be expected during jet propagation \citep[e.g.,][]{MW01,RMW03,RMW04,AB05,HA08,IMNS08}. One may expect TeV--PeV neutrino emission associated with photospheric and/or shock breakout emissions, where the formation of collisionless shocks is expected \citep{M08,MTLB11,KSW11}. At larger dissipation radii above the photosphere, PeV--EeV neutrinos can be produced \citep[e.g.,][]{WB97,DA03,MN06}. Although nuclei also produce neutrinos, the condition that nuclei survive limits the amount of neutrino-generating collisions, so the neutrino flux is typically small and not so easy to detect by IceCube \citep[e.g.,][]{MB10b}. We emphasize that in heavy nuclei scenarios, the GRB--UHECR hypothesis is consistent with the latest upper limits placed by IceCube \citep{Ice12}. Also, it would be hard to distinguish the neutrino signature from those of protons \citep{MINN08}.

The existence of nuclei can also be probed by identifying atomic or nuclear line emissions. As in discussed \cite{MR01}, the cocoon material produced by the jet would break out of the stellar envelope, where iron-enriched clumps may be shed by a UV/X-ray continuum and an Fe line luminosity of $\sim 4 \times 10^{47}~{\rm erg \, s^{-1}}~{(M_b/{10}^{-5} M_{\odot})} x_{\rm Fe}$ is expected. Here $M_{b}$ is the bubble mass and $x_{\rm Fe}$ is the Fe mass fraction. Alternatively, the Fe line emission may be caused by X-rays produced by a continuous but decaying jet \citep{RM00}. Though the details are uncertain, if the high abundance of heavy nuclei is realized, X-ray line features might be constrained or even detected by current and future X-ray satellites such as Astro-H. Furthermore, it is possible that boosted atomic photons could be expected in the GeV range for accelerated nuclei \citep{KV11}. 

Nuclei can also emit $\sim $ MeV gamma rays (in their rest frame) through their excitation states. Such nuclear gamma rays are not so easy to detect for distant GRBs, but they may be detected for nearby and/or luminous events. In particular, nuclear gamma rays may be boosted to the TeV energy range when nuclei are accelerated. For example, the radioactive isotopes $^{56}$Ni and $^{56}$Co lead to the production of MeV nuclear gamma-ray lines after they are excited via, e.g., electron capture or positron emission \citep[for details, see, e.g.,][]{M04,HB10}. These nuclei can be entrained and lead to GeV--TeV gamma rays in GRBs and hypernovae \citep{IM10}. The time scale of emission is also extended by the nuclei Lorentz factor to $\sim {10}^{5}~{\rm yr}~\gamma_{A,5}$ for $^{56}$Co (where $\gamma_A$ is the Lorentz factor of the nuclei in the observer frame), and detection is possible only for Galactic events. On the other hand, nuclei interact with low-energy photons in the source via the photodisintegration process, leading to subsequent (almost prompt) $\sim 0.2~{\rm TeV}~\gamma_{A,5}$ gamma rays from daughter excited nuclei. This signal is useful as a unique probe of nuclei acceleration as well as synchrotron and inverse-Compton gamma rays from pairs generated via the Bethe--Heitler process \citep{MB10a,AT10}. There are a multitude of potential signatures to be explored.

\begin{acknowledgments}   

We thank John Beacom and Brian Metzger for discussions and comments. S.H.~and K.M.~are supported by the Center for Cosmology and Astro-Particle Physics (CCAPP) at the Ohio State University. K.M.~is also supported by JSPS. K.I.~is supported by grants-in-aid from the Ministry of Education, Culture, Sports, Science, and Technology (MEXT) of Japan Nos. 19047004, 22244019, 22244030, 21684014. P.M.~is supported by National Science Foundation grant PHY-0757155. 

\end{acknowledgments}   


\end{document}